\documentclass[%
 reprint,
superscriptaddress,
 amsmath,amssymb,
 aps,
floatfix,
showkeys,
]{revtex4-2}

\usepackage{physics}
\usepackage{mathdots}
\usepackage{graphicx}
\usepackage{dcolumn}
\usepackage{bm}

\newcommand{\CRM}{Centre de recherches mathématiques, Université de Montréal\\
P.O. Box 6128, Centre-ville Station, Montréal (Québec), H3C 3J7, Canada}

\newcommand{\DepPhys}{Département de Physique, Université de Montréal, Montréal (Québec), H3C 3J7, Canada}

\newcommand{\IVADO}{Institut de valorisation des données (IVADO), Montréal (Québec), H2S 3H1, Canada}

\newcommand{\Renmin}{School of Mathematics, Renmin University of China, Beijing, 100872, China}

\begin{document}

\preprint{APS/123-QED}

\title{Analytic ``Newton’s cradles'' with perfect transfer and fractional revival}%

\author{Hugo Schérer}
\email{hugo.scherer@umontreal.ca}
\affiliation{\CRM}
\affiliation{\DepPhys}
\author{Luc Vinet}
\email{luc.vinet@umontreal.ca}
\affiliation{\CRM}
\affiliation{\DepPhys}
\affiliation{\IVADO}
\author{Alexei Zhedanov}
\email{zhedanov@yahoo.com}
\affiliation{\CRM}
\affiliation{\Renmin}

\date{\today}

\begin{abstract}
Analytic mass-spring chains with dispersionless pulse transfer and fractional revival are presented. These are obtained using the properties of the para-Racah polynomials. This provides classical analogs of the quantum spin chains that realize important tasks in quantum information: perfect state transfer and entanglement generation.
\end{abstract}

\maketitle

\section{Introduction}
\label{sec:intro}

Perfect state transfer refers to the transportation of quantum states from one location to another with unit probability. Fractional revival consists in the periodic resurgence of a state at certain sites. Both phenomena are of importance in quantum information. It has been shown how to engineer analytic spin chains or photonic lattices (see \cite{BosseVinet_2017} for a review and references) which dynamically enact these processes. Their design relies on the theory of orthogonal polynomials in the context of inverse spectral problems \cite{Gladwell}.

The examination of classical systems with similar features is certainly worth of interest and it was recently shown \cite{Vaia_NewtonCradle, Vaia_Matrix} that a non-uniform mass-spring chain with perfect transfer could be constructed using what turns out to be the dual Hahn polynomials. We aim to provide here a broader family of such analytic ``Newton's cradles'' and to identify mass-spring chains that exhibit fractional revival (with or without perfect transfer).

As mentioned, orthogonal polynomials play a central role in the construction of such analytic models \cite{Vinet_HowTo, Vinet_2012, Lemay_FR_and_paraRacah, Christandl_2017, Coutinho_2019}. In dealing with quantum systems, a condition on the associated polynomials for perfect transfer to occur is that their spectra be given in terms of integer. It is necessary for the corresponding oscillating system to ever return to its initial state. An additional constraint in the case of classical oscillating systems is that the integers must be perfect squares.

We shall thus study mass-spring chains viewed as collections of $N+1$ masses $\{m_i\}_{i=0}^N$ joined by springs obeying Hooke's law and with elastic constants $\{K_i\}_{i=0}^{N+1}$. They will be said to be free-free ($K_0 = K_{N+1} = 0$), fixed-fixed ($K_0 \neq 0 \neq K_{N+1}$), or fixed-free ($K_0 \neq 0$ and $K_{N+1} = 0$, or vice versa) depending if the first and last masses are joined to a wall by a spring or not. Such systems may be used to describe atomic chains, or oscillating LC circuits when replacing masses and springs by inductors and capacitors. We shall take as the classical version of the quantum perfect state transfer the situation where a pulse given to the first mass is fully transported (i.e. without dispersion) to the last mass in a finite time \cite{Vaia_NewtonCradle}. The issue is then to determine if the masses and spring constants can be chosen (analytically) so that such perfect transfer can be achieved in mass-spring chains.

Analytic blueprints for spin chains and graphs (or optical lattices) with fractional revival, whereby the initial state is replicated on a limited number of sites exclusively, have also been worked out \cite{Lemay_FR_and_paraRacah, Genest_FR, Genest_2016, Christandl_2017}. It has been shown that systems with fractional revival can be obtained in particular from models with perfect state transfer through isospectral deformations. The natural classical equivalent of this quantum phenomenon in the case of mass-spring chains is the distribution of the initial pulse on a restricted number of masses, all other masses having zero momentum. We shall also look for classical chains with this feature.

Our general goal will hence be to obtain analytic free-free and fixed-fixed mass-spring chains that exhibit perfect transfer and fractional revival, in the above sense. This will be achieved by making use of the not-so-well-known para-Racah polynomials, a finite set of $N+1$ polynomials that have a quadratic bi-lattice as a spectrum and which were discovered recently \cite{paraRacah}. These polynomials have already appeared in the construction of perfect quantum spin chains \cite{Lemay_FR_and_paraRacah}. They have different expressions depending on wether $N$ is odd or even, and these cases will be treated separately throughout the paper.

The outline is the following. We first develop the free-free system. In section \ref{sec:freePST}, we describe the problem in detail. In sections \ref{sec:PSTNodd} and \ref{sec:PSTNeven}, we construct the mass-spring chains with the desired properties using the para-Racah polynomials, for $N$ odd and $N$ even respectively. In section \ref{sec:FR_isoDef}, we study fractional revival, looking in particular at isospectral deformations. In section \ref{sec:surgery}, we discuss spectral surgery as a mean to obtain new chains with the desired features. Finally in section \ref{sec:fixed}, we analyse the fixed-fixed system.

\section{Free-free mass-spring chain and perfect transfer}
\label{sec:freePST}

We follow the presentation given in \cite{Vaia_NewtonCradle}. The free-free mass-spring chain is characterized by the values of the $N+1$ masses $\{m_i\}_{i=0}^N$ and the $N$ spring constants $\{K_i\}_{i=1}^N$. With $P_i$ as the momentum of the $i$-th mass and $Q_i$ its displacement from equilibrium, the Hamiltonian of the system is given by:
\begin{equation}
    \mathcal{H} = \sum_{i=0}^N \frac{P_i^2}{2m_i} + \frac{1}{2} \sum_{i=1}^N K_i \qty(Q_{i-1} - Q_i)^2.
\end{equation}

\noindent It is convenient to represent this Hamiltonian in matrix form with $P$ and $Q$ the vectors in $\mathbb{R}^{N+1}$ with components $P_i$ and $Q_i$ respectively. Defining the matrices $M$ and $K$ as follows:
\begin{equation}
    M = \mqty(
    m_0 & 0 &&  \\
    0 & m_1 & 0 &  \\
    & 0 & m_2 & \\
     &  & & \ddots & 0\\
    &&& 0 & m_N
    )_{N+1},
\end{equation}
\begin{equation}
    K = \mqty(
    K_1 & -K_1 & 0 &  \\
    -K_1 & K_1 + K_2 & -K_2 &  \\
    0 & -K_2 & K_2 + K_3 & \ddots\\
     & & \ddots & \ddots & -K_N\\
    &&& -K_N & K_N
    )_{N+1},
\end{equation}

\noindent the Hamiltonian can be written as
\begin{equation}
    \mathcal{H} = \frac{1}{2} P^T M^{-1} P + \frac{1}{2} Q^T K Q.
\end{equation}

\noindent We can furthermore define the mass-weighted coordinates and momenta,
\begin{equation}
    q = M^{\frac{1}{2}}Q, \quad p = M^{-\frac{1}{2}}P,
\end{equation}

\noindent to find
\begin{equation}
    \mathcal{H} = \frac{1}{2} p^T p + \frac{1}{2} q^T A q
    \label{eq:HamA}
\end{equation}

\noindent with $A = M^{-\frac{1}{2}} K M^{-\frac{1}{2}}$, a Jacobi, or tridiagonal symmetric, matrix given by
\begin{equation}
    A = \mqty(
    b_0 & -a_1 & 0 & \\
    -a_1 & b_1 & -a_2 &  \\
    0 & -a_2 & b_2 & \ddots\\
     & & \ddots & \ddots & -a_N\\
    &&& -a_N & b_N
    )_{N+1}
    \label{eq:mat_A}
\end{equation}

\noindent where
\begin{align}
    b_i &= \frac{K_i + K_{i+1}}{m_i}, \quad i = 0,\dots,N,
    \label{eq:bi_matrix}\\
    a_i &= \frac{K_i}{\sqrt{m_{i-1}m_i}}, \quad i = 1,\dots,N,
    \label{eq:ui_matrix}
\end{align}

\noindent with $K_0 = K_{N+1} = 0$ so that the expressions are true for any $n$. Notice that the system is scale invariant, since multiplying all the masses and spring constant by a constant will give the same matrix $A$.

To obtain end-to-end perfect transfer, it is necessary that the chain be mirror-symmetric \cite{Albanese_2004, Kay_2010, Vinet_HowTo}, i.e. that $m_i = m_{N-i}$ and $K_i = K_{N+1-i}$. An immediate consequence is that the matrix $A$ is persymmetric, meaning that it is invariant under reflection with respect to the antidiagonal, i.e. $b_i = b_{N-i}$ and $a_i = a_{N+1-i}$, or equivalentely, $RAR=A$ with
\begin{equation}
    R = \mqty(
    &&& 0 & 1 & \\
    && 0 & 1 & 0  \\
    & & 1 & 0\\
    0 & \iddots &&&\\
    1 & 0 &&&
    )_{N+1}.
\end{equation}

Let $U$ be the matrix that diagonalizes $A$,
\begin{equation}
    UAU^T = D \qq{with} D_{mn} = \delta_{mn} \lambda_n,
\end{equation}

\noindent i.e. the $n$-th line of $U$ is the normalized eigenvector corresponding to eigenvalue $\lambda_n$. Because $A$ is symmetric, $U$ is orthogonal ($UU^T = I$) and the eigenvalues $\lambda_n$ are real. Furthermore, the fact all $a_i$ (in $A$) are nonzero implies that all the eigenvalues $\lambda_n$ are distinct, and the fact that $A$ is positive, semi-definite implies that these eigenvalues are non-negative \cite{Gladwell}. Upon introducing the normal-mode coordinates and momenta,
\begin{equation}
    \tilde{q} = U^T q, \quad \tilde{p} = U^T p,
\end{equation}

\noindent the Hamiltonian becomes that of $N+1$ independent oscillators with ``spring constants'' $\lambda_n$,
\begin{equation}
    \mathcal{H} = \frac{1}{2} \tilde{p}^T \tilde{p} + \tilde{q}^T D \tilde{q}
    = \frac{1}{2} \sum_{n=0}^N \qty(\tilde{p}_n^2 + \lambda_n \tilde{q}_n^2).
\end{equation}

\noindent It is thus manifest that the normal-mode frequencies $\omega_n$ of the system are given by
\begin{equation}
    \lambda_n = \omega_n^2
    \label{eq:xn_wn2}
\end{equation}

We will assume from now on that the $\omega_n$ are ordered, i.e. $\omega_0 < \omega_1 < \dots < \omega_N$. It is obvious that the eigenvalues of $A$ have to be distinct and non-negative and that we must have $\omega_0 = 0$, to account for the fact that the chain is free-free and that there exists a translation mode.

We can describe explicitly the motion of each mass by
\begin{equation}
    q_i(t) = \sum_{n=0}^N U_{ni} \sum_{j=0}^N U_{nj} \qty[q_j(0) \cos \omega_n t + p_j(0) \frac{\sin \omega_n t}{\omega_n}]
\end{equation}
with the understanding that when $n = 0$, $\frac{\sin \omega_n t}{\omega_n} \mapsto t$.

To consider perfect transfer, we take the initial conditions
\begin{equation}
    q(0) = (0,0,\dots,0)^T, \quad p(0) = (\bar{p},0,\dots,0)^T,
    \label{eq:init_cond}
\end{equation}

\noindent which yields
\begin{equation}
    q_i(t) = \bar{p} \sum_{n=0}^N U_{ni} U_{n0} \frac{\sin \omega_n t}{\omega_n}.
\end{equation}

\noindent We are interested in the evolution of the momentum of each mass, which is given by
\begin{equation}
    p_i(t) = \partial_t q_i(t) = \bar{p} \sum_{n=0}^N U_{ni} U_{n0} \cos \omega_n t.
\end{equation}

\noindent Perfect transfer is achieved if there exists a time $t^*$ such that
\begin{equation}
    p(t^*) = (0,0,\dots,0,\bar{p}).
    \label{eq:PST_final}
\end{equation}

\noindent Eigenvectors of $A$ alternate between mirror-symmetric and mirror-antisymmetric vectors \cite{Cantoni1976}, i.e.
\begin{equation}
U_{n,N-i} = (-1)^n U_{ni},
\label{eq:alternate}
\end{equation}

\noindent so we have
\begin{equation}
    \frac{p_N(t)}{\bar{p}} = \sum_{n=0}^N U_{n0}^2 \cos(n\pi - \omega_n t).
\end{equation}

\noindent Perfect transfer will be achieved if $p_N(t)/\bar{p} = 1$, that is if
\begin{equation}
    n\pi - \omega_nt^* = \text{(even integer)} \cp \pi.
\end{equation}

\noindent Equivalentely, this amounts to having
\begin{equation}
    \omega_n = \omega k_n
    \label{eq:wn_wkn}
\end{equation}

\noindent with $\omega = \pi/t^*$ and $k_n$ distinct integers with the same parity as $n$ and no common factor, which is equivalent to
\begin{equation}
    \delta_n = k_{n+1} - k_{n}
\end{equation}

\noindent being odd positive integers with no common factor.

\section{A model with perfect transfer based on the para-Racah polynomials for $N$ odd}
\label{sec:PSTNodd}

The para-Racah polynomials are a finite set of $N+1$, orthogonal polynomials. They arise from a non-standard truncation of the Wilson polynomials \cite{paraRacah}. Let us work first with $N$ odd,
\begin{equation}
    N = 2j + 1, \quad j=0,1,2,\dots \label{eq:Nodd}
\end{equation}

We shall deal with the monic normalization of these polynomials, $\check{P}_n(x^2) = \check{P}_n(x^2; N; a,c,\alpha) = x^{2n} + \dots$ (The use of capital $P$ to designate polynomials or momenta should not lead to confusion as context will always make the intent clear.) As monic orthogonal polynomials, the $\check{P}_n(x^2)$ obey a three-term recurrence relation of the form
\begin{equation}
    x^2 \check{P}_n(x^2) = \check{P}_{n+1}(x^2) + \check{b}_n \check{P}_n(x^2) + \check{u}_n \check{P}_{n-1}(x^2)
    \label{eq:recRel}
\end{equation}

\noindent where the recurrence coefficients given in \cite{paraRacah} depend in this case on the parameters $a$, $c$ and $\alpha$. Up to normalisation, these polynomials provide the lines of the matrix $U$ that diagonalizes the Jacobi matrix $A$ with corresponding entries $b_n = \check{b}_n$ and $a_n = \sqrt{\check{u}_n}$. Moreover, when $\alpha = \frac{1}{2}$, the recurrence coefficients of the para-Racah define a tridiagonal persymmetric matrix. These polynomials are orthogonal on a finite set of points $\lambda_n$ that form a quadratic bi-lattice and are the eigenvalues of $A$,
\begin{align}
    \check{\lambda}_{2s} &= (s+a)^2, \quad s = 0,\dots,j,
    \label{eq:x2sOdd}\\
    \check{\lambda}_{2s+1} &= (s+c)^2, \quad s = 0,\dots,j.
    \label{eq:x2s1Odd}
\end{align}

\noindent The quadratic nature of this spectrum will allow to design a system of masses and springs based on the para-Racah polynomials and exhibiting perfect transmission and fractional revival. (Upon comparing with \cite{paraRacah}, the reader will observe that the signs of the eigenvalues have been reversed and that the recurrence coefficients and polynomials have been transformed accordingly.)

It is in general straightforward to determine the correspondence between the Jacobi matrices and polynomials with spectra that are related by an affine transformation such as
\begin{equation}
    \lambda_n = \Omega (\check{\lambda}_n + \Delta).
    \label{eq:affineTransfo}
\end{equation}

\noindent This goes as follows:
\begin{equation}
    b_n = \Omega (\check{b}_n + \Delta), \quad u_n = \Omega^2 \check{u}_n,
    \label{eq:affine_bn_un}
\end{equation}

\noindent and the new monic polynomials $P_n(x^2)$ with these $b_n$ and $u_n$ as recurrence coefficients are related to the original ones $\check{P}_n(x^2)$ by
\begin{equation}
    P_n(x^2) = \Omega^n \check{P}_n \qty(\frac{x^2}{\Omega} - \Delta).
\end{equation}

We shall use this latitude to appropriately set the parameters for our purpose. If we choose $\Delta = 0$ and $\Omega = \tilde{\omega}^2$ a real, positive parameter, then from (\ref{eq:xn_wn2}), (\ref{eq:x2sOdd}) and (\ref{eq:x2s1Odd}), we will have the following frequencies for the normal modes,
\begin{align}
    \omega_{2s} &= \tilde{\omega} |s+a|, \quad s = 0,\dots,j,\\
    \omega_{2s+1} &= \tilde{\omega} |s+c|, \quad s = 0,\dots,j.
\end{align}

\noindent There is a need for one of the eigenvalues to be zero. Combined with the constraints to respect the positivity of $u_n$ (see \cite{paraRacah}), one convenient choice of parameters is
\begin{align}
    &a=0 \label{eq:aFree},\\
    &0 < c < 1 \label{eq:cFree}.
\end{align}

\noindent It can be shown that any other choice of parameters respecting the constraints in \cite{paraRacah} and having a non-degenerate spectrum with one eigenvalue being zero, will produce the same spectrum and matrix coefficients that $a=0$ and $0<c<1$ yield. No generality will therefore lost by fixing $a$ and $c$ according to (\ref{eq:aFree}) and (\ref{eq:cFree}). Furthermore, the parameter $c$ needs to be a fraction 
\begin{equation}
    c = \frac{\rho}{Z}
    \label{eq:cFreeFraction}
\end{equation}

\noindent with $\rho$ an odd integer, $Z$ an even integer, and $\rho$ and $Z$ co-prime. The eigenfrequencies become
\begin{align}
    \omega_{2s} &= \frac{\tilde{\omega}}{Z} (Zs), \quad s = 0,\dots,j,\\
    \omega_{2s+1} &= \frac{\tilde{\omega}}{Z} (Zs+\rho), \quad s = 0,\dots,j,
\end{align}

\noindent which respects (\ref{eq:wn_wkn}) with $\omega = \tilde{\omega}/Z$. This choice of parameters gives the following entries for the matrix $A$,
https://fr.overleaf.com/project\begin{equation}
    b_n = \frac{\tilde{\omega}^2}{2}\qty[c(c+j)+n(N-n)]
\end{equation}

\noindent and $a_n = \sqrt{u_n}$ with
\begin{align}
    u_n = \tilde{\omega}^4 \frac{n(N+1-n)(N-n+c)(n-1+c)}{4(N-2n)(N-2n+2)} \nonumber\\
    \hfill \vdot \qty((n-j-1)^2-c^2).
\end{align}

\noindent The entries of $U$ are given by the orthonormal polynomials, with $w_n$ the weights with respect to which they are orthogonal \cite{paraRacah},
\begin{equation}
    U_{ni} = \frac{\sqrt{w_n} P_i (\lambda_n)}{\sqrt{u_1 \dots u_i}}.
    \label{eq:Uni}
\end{equation}

To solve for the masses and spring constants, we first define $y_i$,
\begin{equation}
    y_i = \sqrt{\frac{m_{i+1}}{m_i} u_{i+1}} =
    \frac{K_{i+1}}{m_i}.
    \label{eq:yi}
\end{equation}

\noindent From (\ref{eq:bi_matrix}) and (\ref{eq:ui_matrix}), we derive the following recurrence relation for $y_i$,
\begin{align}
    y_i &= b_i - \frac{u_i}{y_{i-1}},\\
    y_0 &= b_0.
\end{align}

\noindent In obtaining the para-Racah polynomials from the Wilson polynomials \cite{paraRacah}, the following relations are used to define and calculate $b_n$ and $u_n$,
\begin{align}
    b_n &= A_n + C_n + a^2, \label{eq:b_nFromAC}\\
    u_n &= A_{n-1} C_n,
\end{align}

\noindent with
\begin{align}
    A_n &= \tilde{\omega}^2 \frac{(N-n)(n+c)(n-c-j)}{2(2n-N)},\\
    C_n &= \tilde{\omega}^2 \frac{n(N-n+c)(n-j-1+c)}{2(2n-N)}.
\end{align}

\noindent It is easy to see that, when $a=0$ as in our case, $A_i$ obeys the same recurrence relation as $y_i$, and that $A_0 = b_0$ since $C_0 = 0$. We conclude that
\begin{equation}
    y_i = A_i.
\end{equation}

\noindent The formulas for $m_i$ and $K_i$ can now be easily obtained from (\ref{eq:yi}), with $m_0$ as the scaling parameter,
\begin{align}
    m_i &= \frac{(A_{i-1} A_{i-2} \dots A_0)^2}{u_i u_{i-1} \dots u_1} m_0, \label{eq:mi_Aiui}\\
    K_i &= A_{i-1} m_{i-1} \label{eq:Ki_Aiui}.
\end{align}

\noindent From this, we can express the masses and spring constants in closed-forms, 
\begin{align}
    \frac{m_i}{m_0} &= \frac{(-N)_i (c)_i (-j-c)_i}{i! (-N+1-c)_i (-j+c)_i} 
    \vdot \frac{(N-2i)}{N},
    \label{eq:mi_odd}\\
    \frac{K_i}{\tilde{\omega}^2 m_0} &= \frac{i(N-i+c)(i-j-1+c)}{2 (2i-N)} \qty(\frac{m_i}{m_0}),
    \label{eq:Ki_odd}
\end{align}

\noindent in terms of the standard Pochhammer symbol 
\begin{equation}
    (a)_k = a (a+1) (a+2) \dots (a+k-1).
\end{equation}

\noindent Examples of values of the masses and spring constants forming the chain for certain choices of parameters are displayed in figure \ref{fig:free_mK}.

\section{The case of $N$ even}
\label{sec:PSTNeven}

We now turn to $N$ even,
\begin{equation}
    N=2j, \quad j=0,1,2,\dots \label{eq:Neven}
\end{equation}

\noindent Again, we need $\alpha = \frac{1}{2}$ to have a persymmetric matrix. The spectrum is nearly identical to the odd case, with only one less eigenvalue.
\begin{align}
    \lambda_{2s} &= (s+a)^2, \quad s = 0,\dots,j,\\
    \lambda_{2s+1} &= (s+c)^2, \quad s = 0,\dots,j-1.
\end{align}

\noindent The positivity constraints for $u_n$ are different, but combined with the need for one eigenvalue being zero, they lead, again without loss of generality, to the same choice of parameters for $a$ and $c$, and to essentially the same eigenfrequencies for the system
\begin{align}
    \omega_{2s} &= \frac{\tilde{\omega}}{Z} (Zs), \quad s = 0,\dots,j,\\
    \omega_{2s+1} &= \frac{\tilde{\omega}}{Z} (Zs+\rho), \quad s = 0,\dots,j-1,
\end{align}

\noindent with $\omega = \tilde{\omega}/Z$ again. The expressions for $b_n$ and $a_n = \sqrt{u_n}$ are now given by \cite{paraRacah}
\begin{align}
    b_n = \tilde{\omega}^2 \frac{(n-N)(n+c)(n-c-j+1)}{2(2n+1-N)} \nonumber \\
    \hfill + \tilde{\omega}^2 \frac{n(n-N-c)(n-j-1+c)}{2(2n-1-N)},
\end{align}
\begin{align}
    u_n = \tilde{\omega}^4 \frac{n(N+1-n)(c+n-1)(-c+j-n+1)}{4(N-2n+1)^2} \nonumber \\
    \hfill \vdot (c+j-n)(c+N-n),
\end{align}

\noindent and the expression for $A_n$ is
\begin{align}
A_n &= \tilde{\omega}^2 \frac{(N-n)(n+c)(n-c-j+1)}{2(2n+1-N)}.
\end{align}

\noindent The recurrence relation for $y_i$ and $A_i$ is still the same as for the odd $N$ case. Using this, we can obtain closed-form expressions for $m_i$ and $K_i$:
\begin{align}
    \frac{m_i}{m_0} &= \frac{(-N)_i (c)_i (-j-c+1)_i}{i! (-N+1-c)_i (-j+c)_i}, \\
    \frac{K_i}{\tilde{\omega}^2 m_0} &= \frac{i (N-i+c) (i-j-1+c)}{2 (2i-1-N)} \qty(\frac{m_i}{m_0}).
\end{align}

\noindent These formulas when $N$ is even, or the equations (\ref{eq:mi_odd}) and (\ref{eq:Ki_odd}) when $N$ is odd, provide the masses and spring constants of chains with pervect transfer. These models are characterized by the factor $c$ between 0 and 1.

As a special case, we can take $\rho = 1, Z = 2$, so that $c = \frac{1}{2}$ and $\omega = \tilde{\omega}/2$. This reduces to the analytic solution given in \cite{Vaia_NewtonCradle} with $k_n = n$ for both the odd and even cases,
\begin{align}
    b_n &= \frac{1}{2} \qty(\frac{\tilde{\omega}}{2})^2 \qty[N + 4n(N-n)],\\
    \sqrt{u_n} &= \frac{1}{2} \qty(\frac{\tilde{\omega}}{2})^2 \sqrt{n(2n-1)(N+1-n)(2N-2n+1)},
\end{align}

\noindent and
\begin{align}
    m_i &= \binom{N}{i}^2 \binom{2N}{2i}^{-1} m_0,\\
    K_i &= \qty(\frac{\tilde{\omega}}{2})^2 N^2 \binom{N-1}{i-1}^2 \binom{2N}{2i-1}^{-1} m_0.
\end{align}

\begin{figure}[t]
\includegraphics[width=0.45\textwidth]{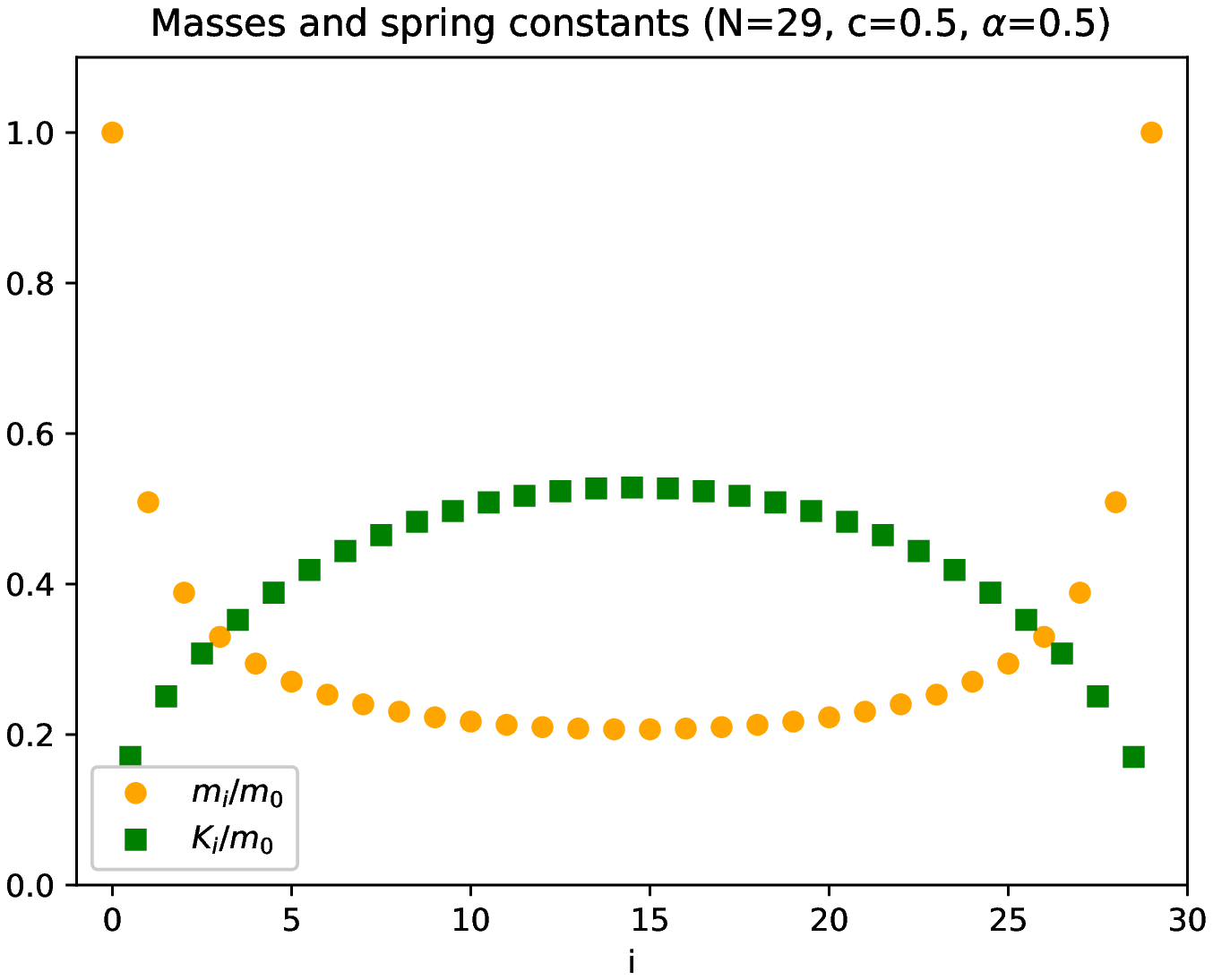}
\includegraphics[width=0.45\textwidth]{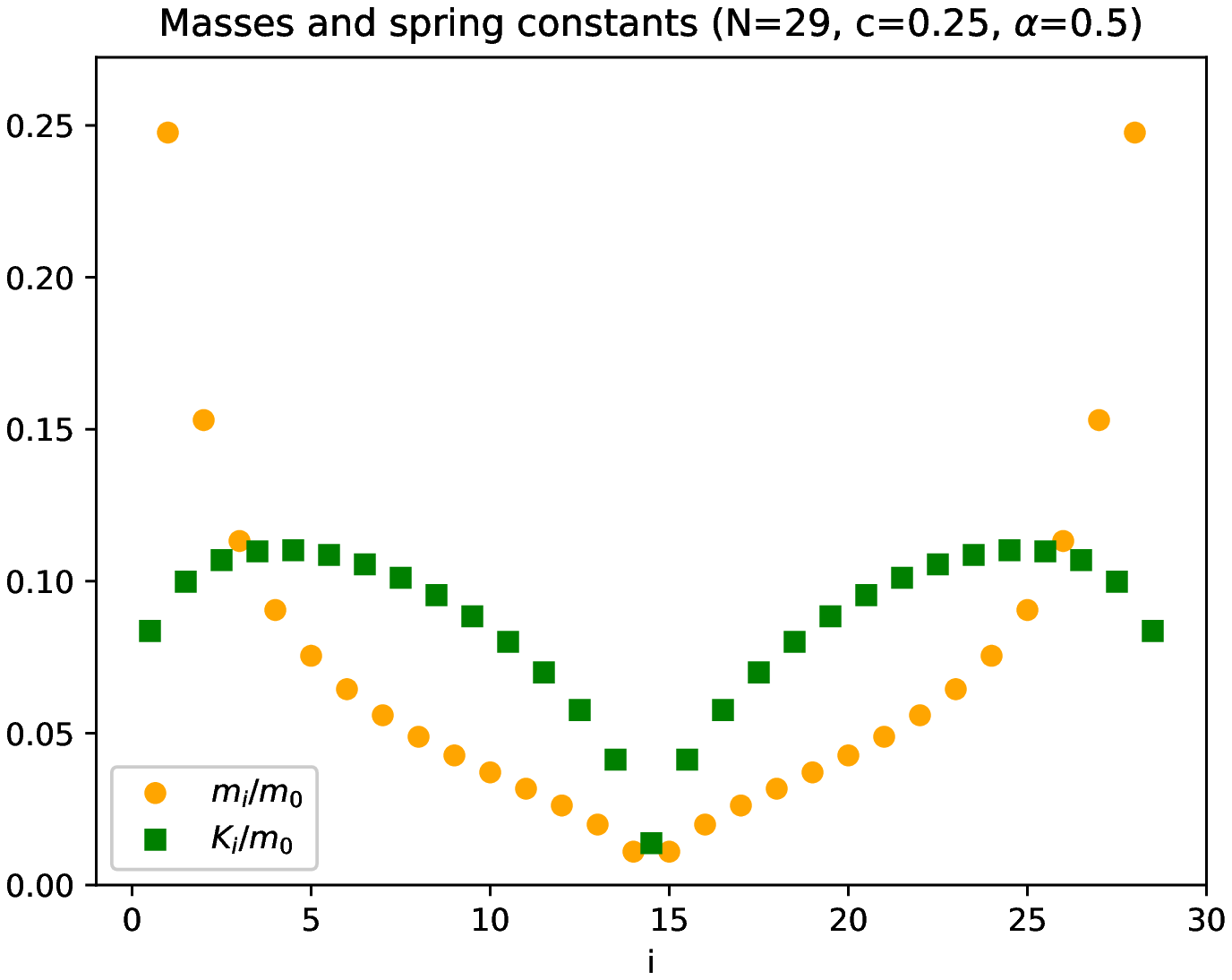}
\includegraphics[width=0.45\textwidth]{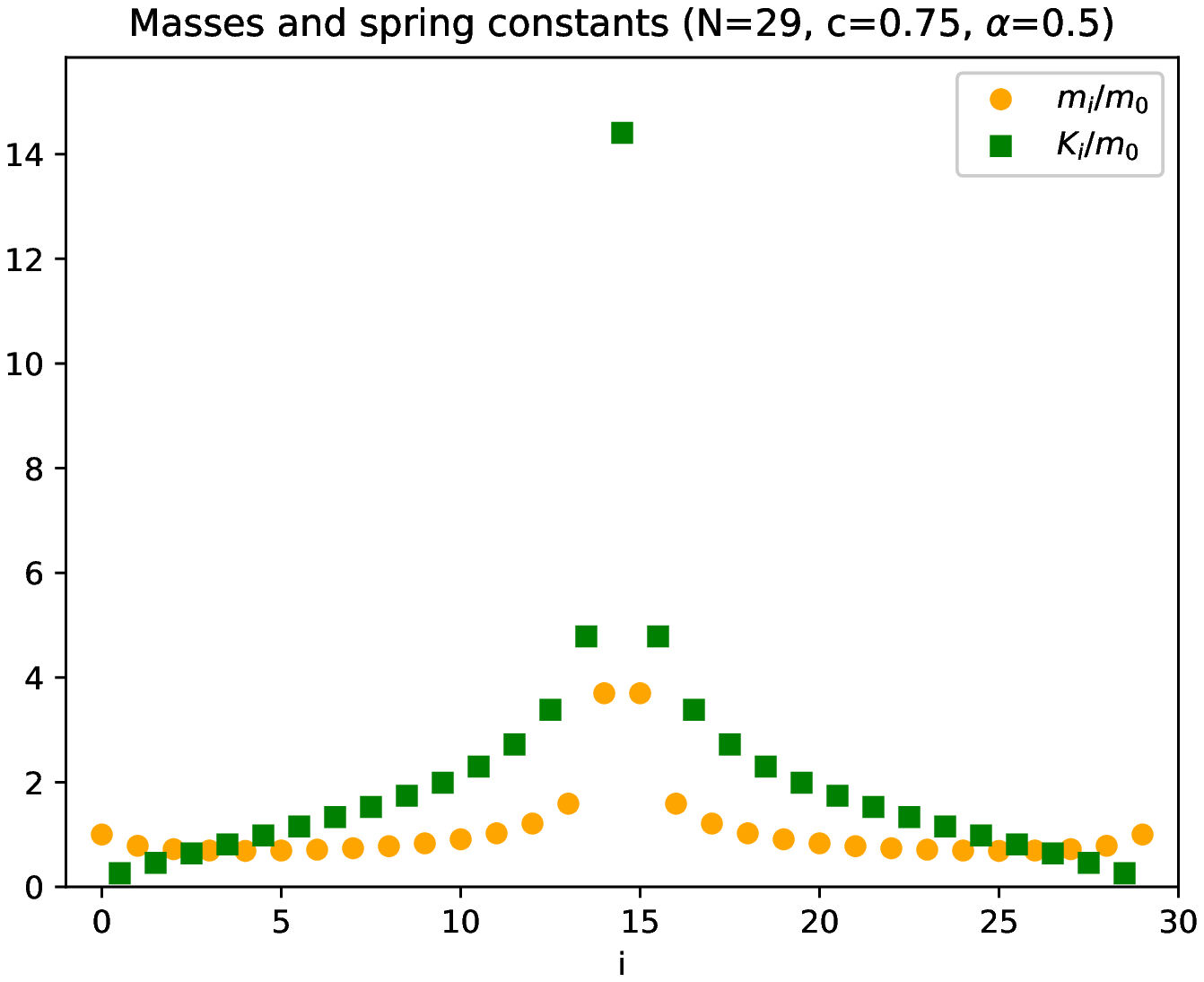}
\caption{\label{fig:free_mK} Normalized masses and spring constants for a mirror-symmetric free-free system with $c=\frac{1}{2}$, $c<\frac{1}{2}$ and $c>\frac{1}{2}$ respectively; with $\tilde{\omega}=2\pi/N$ to allow for adequate visualisation of both quantities in the same graph \cite{Vaia_Matrix}.}
\end{figure}

\begin{figure*}
    \centering
    \includegraphics[width=0.45\textwidth]{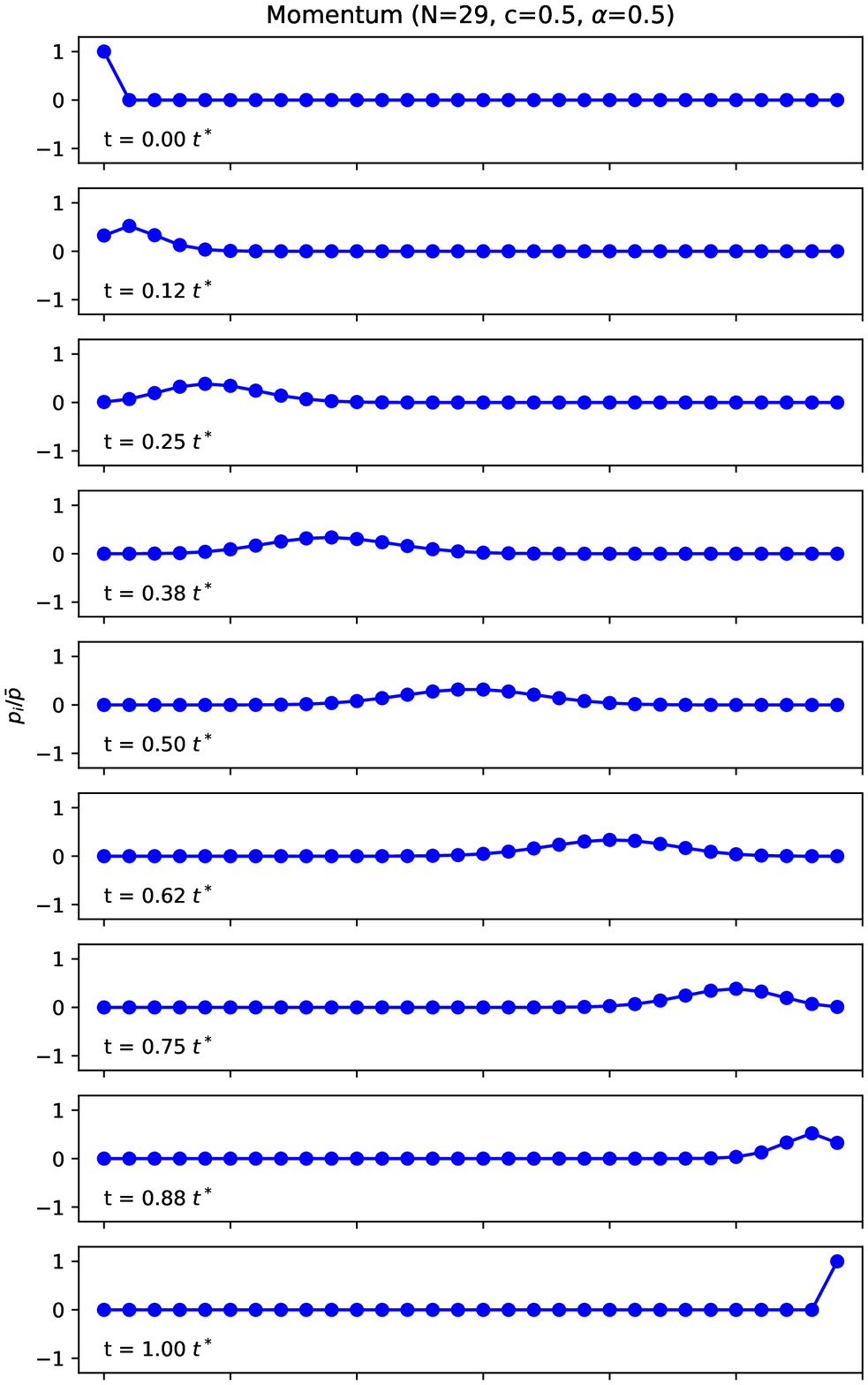}
    \includegraphics[width=0.45\textwidth]{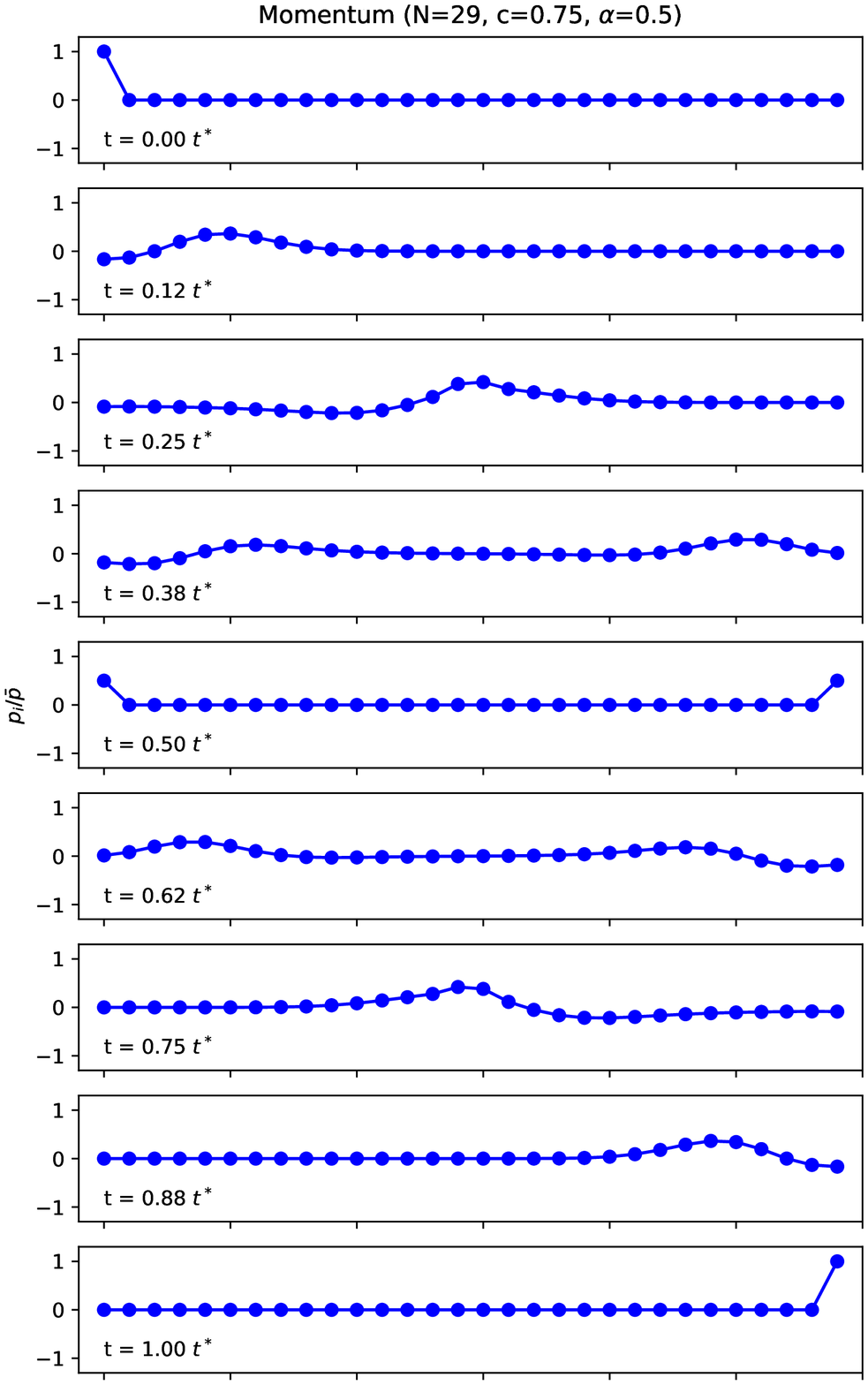}
    \caption{\label{fig:free_p_9}Snapshots of the time-evolution of the mass-weighed momentum of a mirror-symmetric free-free system at various fractions of $t^*$ for $c=\frac{1}{2}$ and $c \neq \frac{1}{2}$ respectively (video available \cite{suppl}).}
\end{figure*}

\noindent Note that this choice of parameters means that we have $c = a + \frac{1}{2}$, and in this case, the para-Racah polynomials are actually orthogonal on a single quadratic lattice and reduce to the dual-Hahn polynomials \cite{paraRacah}. It is interesting to observe that when $c = \frac{1}{2}$, the chain is ``smooth'' in a sense, but when $c \neq \frac{1}{2}$, there is some sort of discontinuity in the middle for the masses and spring constants. When $c<\frac{1}{2}$, the middle masses and spring constants tend to become smaller, whereas when $c>\frac{1}{2}$ they tend to become bigger, this is depicted in figure \ref{fig:free_mK}. Observing the time evolution of the chain in simulations, we see that the momentum wave hits this ``impurity'' in the middle, and is reflected as well as transmitted. Both waves are reflected again at each end of the chain and come back to the middle at the same time, and perfect transfer is subsequently realized after this happens a number of times, see for example figure \ref{fig:free_p_9}.

We now comment on limit cases for $c$. First, when $c \to 0$ we have that $m_0 = m_N$ is a finite positive number (that can be chosen at will since $m_0$ is the scaling parameter), and $m_i = 0$ for all other $i$. This chain actually degenerates into a system of two uncoupled masses free to move. Second, in the limit $c \to 1$, we have that $m_j, m_{j+1}, J_{j+1} \to \infty$ for $N$ odd, and $m_j \to \infty$ for $N$ even. This is as if there is a solid wall at the center of the chain and each half is an independent chain attached to each side of the wall so that no impulsion can be transferred between the two chains anymore.

\section{Fractional revival and isospectral deformation}
\label{sec:FR_isoDef}

Fractional revival in the case of the classical system is a situation where the momentum is strictly shared in a periodic fashion by a limited number of determined masses on the chain, all other masses having zero momentum. Remarkably, fractional revival is observed in the Newton's Cradle based on the para-Racah polynomials. Indeed, it is found that there are certain times during the evolution of this symmetric system when the momentum is entirely distributed on the first and the last masses only.

Let $\tau_\ell = (2\ell/Z) t^*$, $\ell = 0, \dots, Z/2$, with $\tau_0 = 0$ being the initial conditions, and $\tau_{Z/2} = t^*$. Then, 
\begin{align}
    \cos(\omega_{2s} \tau_\ell) &= 1,
    \label{eq:cos2sfree}\\
    \cos(\omega_{2s+1} \tau_\ell) &= \cos(2\ell c\pi).
    \label{eq:cos2s1free}
\end{align}

\noindent These expressions do not depend on $n$ (or $s$) anymore. Furthermore, it can be proved using (\ref{eq:alternate}) that
\begin{align}
    \sum_{s=0}^j U_{2s,i} U_{2s,k} = \frac{1}{2} (\delta_{ik} + \delta_{i,N-k}),
    \label{eq:sumU2s}\\
    \sum_{s=0}^j U_{2s+1,i} U_{2s+1,k} = \frac{1}{2} (\delta_{ik} - \delta_{i,N-k}),
    \label{eq:sumU2s1}
\end{align}

\noindent for $N$ odd, and that the same is true for $N$ even but with the sum in (\ref{eq:sumU2s1}) terminating at $j-1$. Using all these identities, it is straightforward to show that
\begin{equation}
    \frac{p_i(\tau_l)}{\bar{p}} = \delta_{i0} \cos^2(lc\pi) + \delta_{iN} \sin^2(lc\pi).
    \label{eq:pFRfree}
\end{equation}

\noindent This shows that fractional revival occurs on the first and last masses in the mirror-symmetric chain ($\alpha = \frac{1}{2}$) based on the para-Racah polynomials at times $0 \neq \tau_\ell \neq t^*$, when the paramater $c \neq \frac{1}{2}$. Indeed it should be stressed that, as per the definition of $\tau_\ell$, fractional revival can only occur if $Z > 2$ or equivalently if $c \neq \frac{1}{2}$. It is the discontinuity present in the middle of the chain in such cases and the reflection as well as the transmission of the pulse it causes, that ultimately generates fractional revival. This implies that the model presented in \cite{Vaia_NewtonCradle} that has $c=\frac{1}{2}$ does not exhibit fractional revival. As an example, fractional revival can be observed at time $\frac{1}{2}t^*$ on the right side of figure \ref{fig:free_p_9}, but nowhere in the left side where $c=\frac{1}{2}$.

\begin{figure*}
    \centering
    \includegraphics[width=0.45\textwidth]{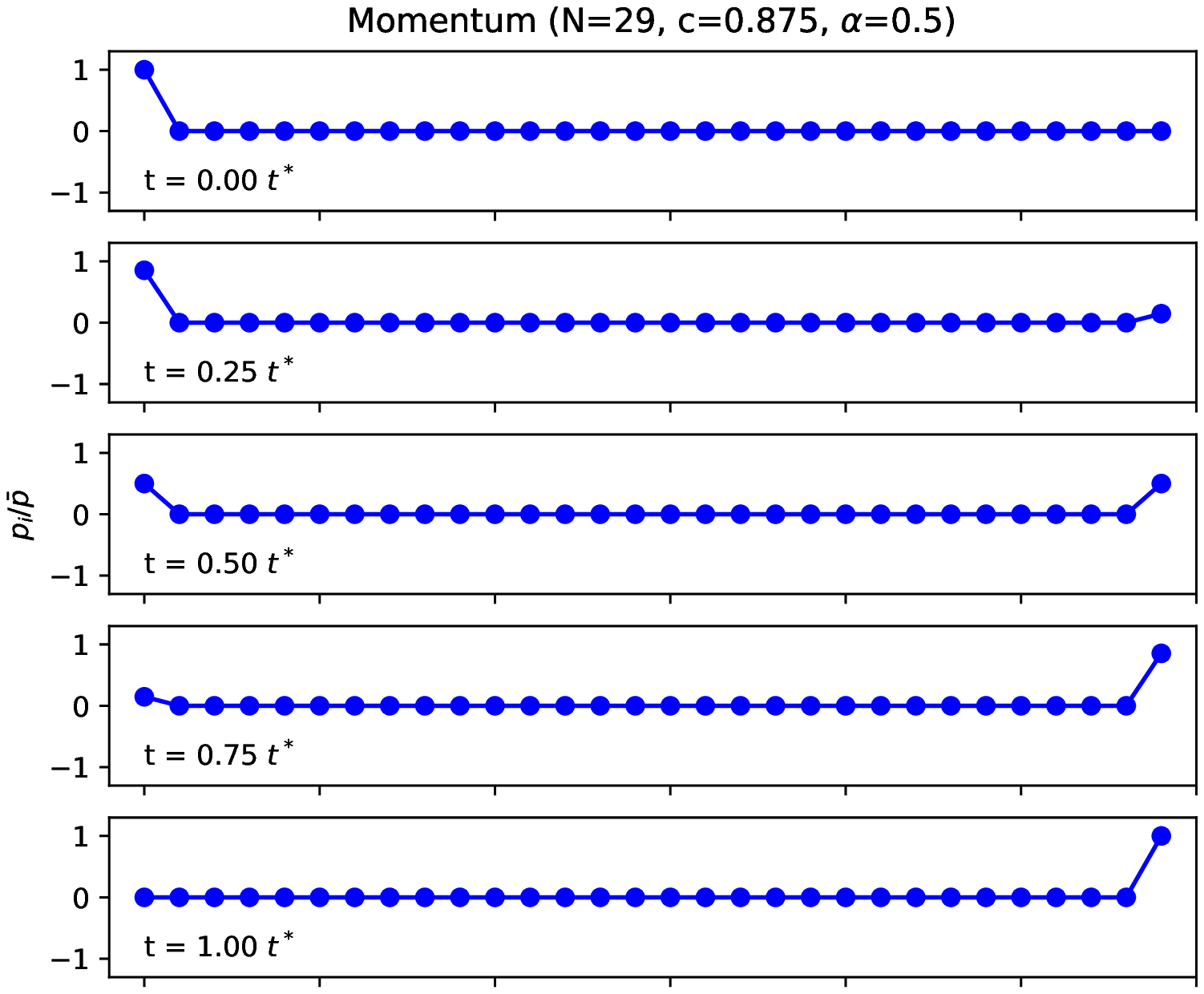}
    \includegraphics[width=0.45\textwidth]{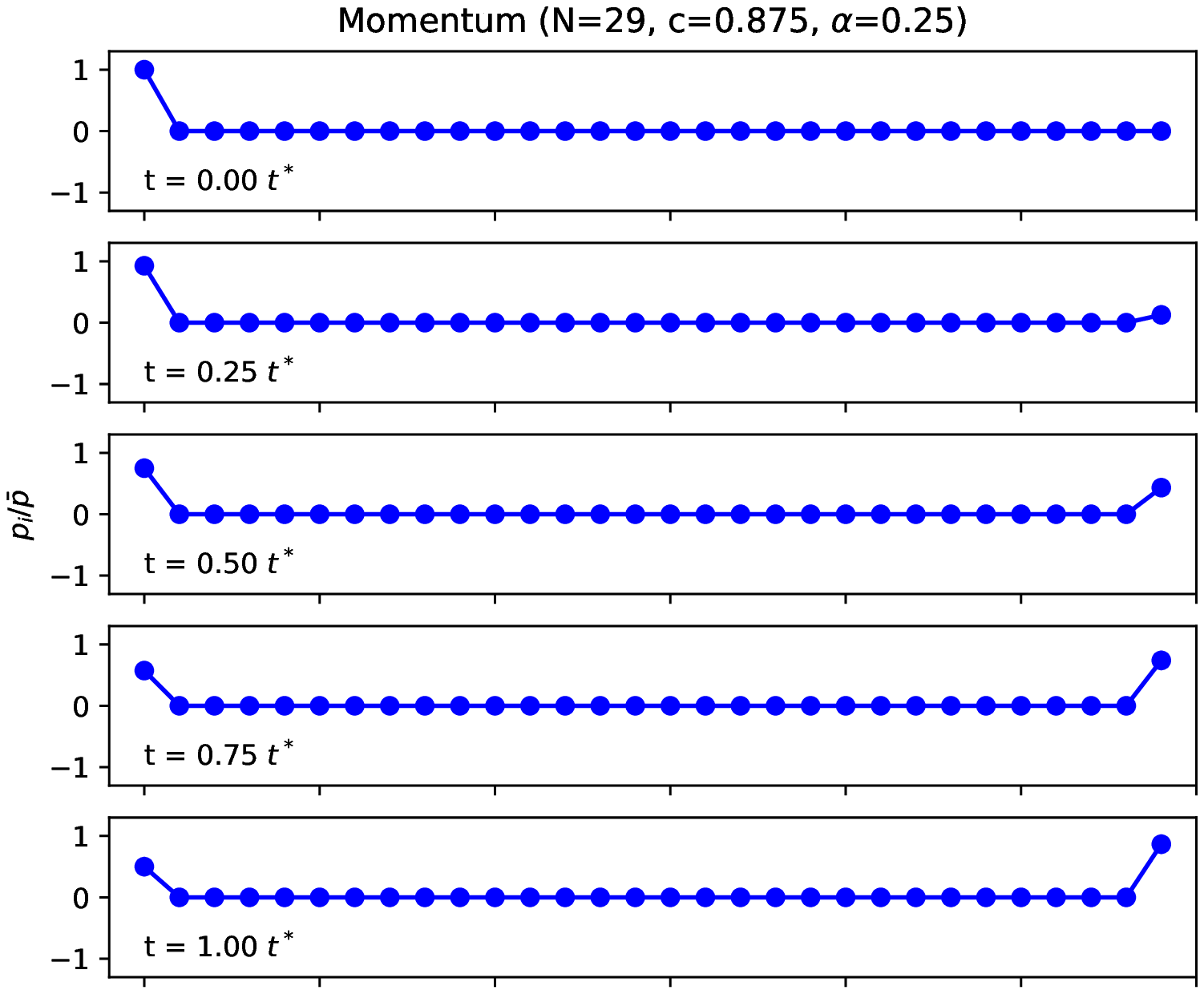}
    \caption{\label{fig:free_p_5}Snapshots of the time-evolution of the mass-weighed momentum of a free-free system at various times $\tau_\ell$ exhibiting fractional revival for $\alpha=\frac{1}{2}$ and $\alpha \neq \frac{1}{2}$ respectively; here $Z=8$ (video available \cite{suppl}).}
\end{figure*}

We can also obtain mass-springs systems with fractional revival using isospectral deformations \cite{Lemay_FR_and_paraRacah, Genest_persymmetric}. Let us consider the matrix $V$,
\begin{equation}
V = \mqty(
    \sin \theta &&&&& \cos \theta \\
    & \ddots &&& \iddots &  \\
    && \sin \theta & \cos \theta && \\
    && \cos \theta & - \sin \theta && \\
    & \iddots &&& \ddots & \\
    \cos \theta &&&&& - \sin \theta
    )_{N+1}
\end{equation}

\noindent for $N$ odd, and
\begin{equation}
V = \mqty(
    \sin \theta &&&&&& \cos \theta \\
    & \ddots &&&& \iddots &  \\
    && \sin \theta & 0 & \cos \theta && \\
    && 0 & 1 & 0 && \\
    && \cos \theta & 0 & - \sin \theta && \\
    & \iddots &&&& \ddots & \\
    \cos \theta &&&&&& - \sin \theta
    )_{N+1}
\end{equation}

\noindent for $N$ even. We see that $V = V^T$ and that $V^2 = I$.

Define $\tilde{A} = VAV$. From here on, symbols with tilde will be associated with the system described by this new matrix $\tilde{A}$, and the symbols with no tilde will correspond to expressions derived from the matrix $A$ with $\alpha = \frac{1}{2}$.

First, observe that the matrix $\tilde{A}$ will have the same spectrum as $A$. Furthermore, only a few entries in the matrix will change. In fact, $\tilde{b}_i = b_i$ and $\tilde{u}_i = u_i$, for all $i$ except
\begin{align}
    \tilde{u}_{j+1} &= u_{j+1} \cos^2(2\theta),\\
    \tilde{b}_j &= b_j + \sqrt{u_{j+1}} \sin(2\theta),\\
    \tilde{b}_{j+1} &= b_j - \sqrt{u_{j+1}} \sin(2\theta),
\end{align}

\noindent for $N$ odd, and
\begin{align}
    \tilde{u}_j = u_j (\cos \theta + \sin \theta)^2,\\
    \tilde{u}_{j+1} = u_j (\cos \theta - \sin \theta)^2,
\end{align}

\noindent for $N$ even. If we relate $\theta$ to $\alpha$ as follows,
\begin{align}
    \sin(2\theta) &= 1 - 2 \alpha,\\
    \cos(2\theta) &= 2 \sqrt{\alpha(1 - \alpha)},
\end{align}

\noindent or, equivalently,
\begin{align}
    \sin \theta &= \frac{\sqrt{1-\alpha} - \sqrt{\alpha}}{\sqrt{2}},\\
    \cos \theta &= \frac{\sqrt{1-\alpha} + \sqrt{\alpha}}{\sqrt{2}},
\end{align}

\noindent the new coefficients $\tilde{b}_i$ and $\tilde{u}_i$ that result are exactly those of the para-Racah polynomials with a general parameter $\alpha$, in the range $0 < \alpha < 1$ as required. Notice how choosing $\alpha = \frac{1}{2}$ will lead to $V=R$, and $\tilde{A} = A$, because $A$ is persymmetric. In terms of $\alpha$, the new entries can be written as
\begin{align}
    \tilde{u}_{j+1} &= 4 \alpha (1-\alpha) u_{j+1},
    \label{eq:tilde_uj1Odd}\\
    \tilde{b}_j &= b_j + (1-2\alpha) \sqrt{u_{j+1}},
    \label{eq:tilde_bjOdd}\\
    \tilde{b}_{j+1} &= b_j - (1-2\alpha) \sqrt{u_{j+1}},
    \label{eq:tilde_bj1Odd}
\end{align}

\noindent for $N$ odd, and
\begin{align}
    \tilde{u}_j = 2(1-\alpha) u_j,
    \label{eq:tilde_ujEven}\\
    \tilde{u}_{j+1} = 2\alpha u_j,
    \label{eq:tilde_uj1Even}
\end{align}

\noindent for $N$ even. In finding expressions (\ref{eq:mi_Aiui}) and (\ref{eq:Ki_Aiui}) for the masses and spring constants, it was not necessary that $\alpha = \frac{1}{2}$, all that was required was that $a=0$. From the para-Racah polynomials, we have $\tilde{A}_i = A_i$ for all $i$ except $i=j$ with $j$ given by (\ref{eq:Nodd}) or (\ref{eq:Neven}), in which case
\begin{equation}
    \tilde{A}_j = 2 \alpha A_j
\end{equation}

\noindent for $N$ odd or even. Using that and (\ref{eq:tilde_uj1Odd}),(\ref{eq:tilde_ujEven}) and (\ref{eq:tilde_uj1Even}), we have
\begin{equation}
    \tilde{m}_i = \begin{cases}
    m_i \qq{if} i \leq j\\
    \frac{\alpha}{1-\alpha} m_i \qq{if} i > j
    \end{cases},
    \label{eq:mi_alpha_odd}
\end{equation}
\begin{equation}
    \tilde{K}_i = \begin{cases}
    K_i \qq{if} i \leq j\\
    2 \alpha K_i \qq{if} i = j+1\\
    \frac{\alpha}{1-\alpha} K_i \qq{if} i > j+1
    \end{cases},
    \label{eq:Ki_alpha_odd}
\end{equation}

\noindent for $N$ odd, and
\begin{equation}
    \tilde{m}_i = \begin{cases}
    m_i \qq{if} i \leq j-1\\
    \frac{1}{2(1-\alpha)} m_i \qq{if} i = j\\
    \frac{\alpha}{1-\alpha} m_i \qq{if} i > j
    \end{cases},
    \label{eq:mi_alpha_even}
\end{equation}
\begin{equation}
    \tilde{K}_i = \begin{cases}
    K_i \qq{if} i \leq j\\
    \frac{\alpha}{1-\alpha} K_i \qq{if} i > j
    \end{cases},
    \label{eq:Ki_alpha_even}
\end{equation}

\noindent for $N$ even. It is interesting and surprising to see that, even though the isospectral deformation only transforms a very limited number of entries in the matrix $A$, it really affects half of the chain when we look at the masses and spring constants. This contrasts with perfect transfer in quantum spin chains where the couplings between sites and magnetic fields on each site are directly the entries of the matrix $A$ and where isospectral deformation only transform the middle of the chain, see for example \cite{Genest_FR, Lemay_FR_and_paraRacah}. It is noteworthy that if (\ref{eq:mi_alpha_odd}) to (\ref{eq:Ki_alpha_even}) are renormalized by choosing $\tilde{m}_0 = 2(1-\alpha) m_0$, effectively multiplying all these equations by $2(1-\alpha)$, one obtains a more symmetrical presentation of the isospectral deformation, where the whole chain is affected.

Clearly, $\tilde{A}$ is diagonalized by $\tilde{U} = UV$. Indeed, 
\begin{equation}
    UV \tilde{A} V^T U^T = U A U^T = D.
\end{equation}

\noindent We recall that if $A$ is diagonalized by $U$, it is also diagonalized by the matrix with entries $(-1)^n U_{ni}$. We shall use this, along with (\ref{eq:alternate}), to find consistent expressions for $\tilde{U}_{ni}$ when $\alpha = \frac{1}{2}$. The new diagonalizing matrix is thus
\begin{equation}
    \tilde{U}_{ni} = \begin{cases}
    U_{ni} \cos \theta + U_{n,N-i} \sin \theta \hfill \qq{if} i \leq j\\
    U_{ni} \cos \theta - U_{n,N-i} \sin \theta \hfill \qq{if} i > j
    \end{cases}
\end{equation}

\noindent for $N$ odd, and 
\begin{equation}
    \tilde{U}_{ni} = \begin{cases}
    U_{ni} \cos \theta + U_{n,N-i} \sin \theta \hfill \qq{if} i < j\\
    U_{ni} \hfill \qq{if} i = j\\
    U_{ni} \cos \theta - U_{n,N-i} \sin \theta \hfill \qq{if} i > j
    \end{cases}
\end{equation}

\noindent for $N$ even. In terms of $\alpha$, with (\ref{eq:alternate}) this gives
\begin{equation}
    \tilde{U}_{ni} = U_{ni} \vdot \begin{cases}
    \sqrt{1+(-1)^n(1-2\alpha)} \hfill \qq{if} i \leq j\\
    \sqrt{1-(-1)^n(1-2\alpha)} \hfill \qq{if} i > j
    \end{cases}
\end{equation}

\noindent for $N$ odd, and
\begin{equation}
    \tilde{U}_{ni} = U_{ni} \vdot \begin{cases}
    \sqrt{1+(-1)^n(1-2\alpha)} \hfill \qq{if} i < j\\
    1 \hfill \qq{if} i=j\\
    \sqrt{1-(-1)^n(1-2\alpha)} \hfill \qq{if} i > j
    \end{cases}
\end{equation}

\noindent for $N$ even. The momentum becomes
\begin{equation}
    \frac{p_i(t)}{\bar{p}} = \begin{cases}
    \displaystyle \sum_{n=0}^N (1+(-1)^n (1-2\alpha)) U_{ni} U_{n0} \cos (\omega_n t)\\ \hfill \qq{if} i \leq j\\
    \displaystyle \sum_{n=0}^N 2\sqrt{\alpha(1-\alpha)} U_{ni} U_{n0} \cos (\omega_n t)\\ \hfill \qq{if} i > j
    \end{cases}
\end{equation}

\noindent for $N$ odd, and
\begin{equation}
    \frac{p_i(t)}{\bar{p}} = \begin{cases}
    \displaystyle \sum_{n=0}^N (1+(-1)^n (1-2\alpha)) U_{ni} U_{n0} \cos (\omega_n t)\\ \hfill \qq{if} i<j\\
    \displaystyle \sum_{n=0}^N \sqrt{(1+(-1)^n (1-2\alpha))} U_{ni} U_{n0} \cos (\omega_n t)\\ \hfill \qq{if} i=j\\
    \displaystyle \sum_{n=0}^N 2\sqrt{\alpha(1-\alpha)} U_{ni} U_{n0} \cos (\omega_n t)\\ \hfill \qq{if} i > j
    \end{cases}
\end{equation}

\noindent for $N$ even. With this, and with equations (\ref{eq:cos2sfree}) to (\ref{eq:sumU2s1}), the momentum at $\tau_l$ can be calculated to be
\begin{align}
    \frac{p_i(\tau_\ell)}{\bar{p}} &= \delta_{i0}(1-2\alpha \sin^2(\ell c\pi)) \nonumber\\
    & + \delta_{iN} (2\sqrt{\alpha(1-\alpha)} \sin^2(\ell c\pi))
    \label{eq:pFRfreeAlpha}
\end{align}

\noindent so fractional revival is also observed at the times $\tau_\ell$, see for example figure \ref{fig:free_p_5}. This reduces to (\ref{eq:pFRfree}) when $\alpha = \frac{1}{2}$. In particular, at time $\tau_{Z/2} = t^*$,
\begin{equation}
    \frac{p_i(t^*)}{\bar{p}} = \delta_{i0}(1-2\alpha) + \delta_{iN} (2\sqrt{\alpha(1-\alpha)}).
\end{equation}

\noindent It should be stressed that perfect transfer cannot be achieved if $\alpha \neq \frac{1}{2}$, but that fractional revival can. We have seen that performing the isospectral transformation on the chain with $\alpha = \frac{1}{2}$ yields the mass-spring system corresponding to the para-Racah polynomials with a generic $\alpha$. This chain is no longer mirror-symmetric, but will exhibit fractional revival (instead of perfect transfer) at time $t^*$. In particular, transforming in that way the chain with $c=\frac{1}{2}$ based on the dual-Hahn polynomials (and presented in \cite{Vaia_NewtonCradle}) will give a system with fractional revival. As for the model with $\alpha = \frac{1}{2}$, resurgences also take place at the times $\tau_\ell \neq t^*$ with the amplitude of the momentum at masses $m_0$ and $m_N$ now modulated by the choice of $\alpha$. Furthermore, looking at (\ref{eq:pFRfreeAlpha}) (and even at (\ref{eq:pFRfree})), one observes that the conditions on the parity of $Z$ and $\rho$ can be relaxed if one only looks for fractional revival and not perfect transfer. One can even choose $c$ to actually be any real number between 0 and 1, noting that the system will however not return to its initial state if $c$ is not a rational number. In summary, the case $\alpha = c = \frac{1}{2}$ leads to a system with perfect transfer only. The case $\alpha = \frac{1}{2}$ and $c \neq \frac{1}{2}$ defined as in (\ref{eq:cFreeFraction}) leads to perfect transfer and fractional revival. With other choices of $c$ (an irreducible fraction that does not respect the parity condition on $\rho$ or $Z$, or an irrational number), the system exhibits fractional revival but no perfect transfer whatever the choice for $\alpha$ is. And when $\alpha \neq \frac{1}{2}$, in this case also, the system only exhibits fractional revival, whatever the choice for $c$ is.

Note that in order to check that momentum is conserved, we need to return to the true momentum $P_i$,
\begin{equation}
    P_i = \sqrt{\tilde{m}_i} p_i.
\end{equation}

\noindent One observes for instance that the momentum at $\tau_l$ is
\begin{align}
    \frac{P_i(\tau_\ell)}{\bar{p}} &= \delta_{i0}(1-2\alpha \sin^2(\ell c\pi))\sqrt{m_0} \nonumber\\
    &+ \delta_{iN} (2\alpha \sin^2(\ell c\pi)) \sqrt{m_0}
\end{align}

\noindent and readily sees that it is conserved.

The case of $\alpha = 0$ is special and degenerate. Indeed, from equations (\ref{eq:mi_alpha_odd}) to (\ref{eq:Ki_alpha_even}), we readily see that the second part of the chain vanishes. We can also multiply all these equation by $\frac{1-\alpha}{\alpha}$ before setting $\alpha = 1$ to see that the first half of the chain then vanishes. The transmission of the pulse is now impossible, as there is no final mass to receive it or no initial mass to give it to.

\section{Spectral surgery}
\label{sec:surgery}

A procedure called spectral surgery developped in \cite{Vinet_HowTo} gives a method to obtain a new matrix $\check{A}$ with perfect transfer from a matrix $A$ with this feature. It relies on the Christoffel transform of the weights $w_s$:
\begin{equation}
    \check{w}_s = C (\lambda_s - \lambda_k) w_s, \quad s = 0,1,\dots,k-1,k+1,\dots,N
    \label{eq:weightSurg}
\end{equation}

\noindent where $C$ is a normalisation constant, and the use of the polynomials $\check{P}_n(x^2)$ which are orthogonal with respect to $w_s$. These polynomials are obtained from the original ones by the formulas
\begin{equation}
    \check{P}_n(x^2) = \frac{P_{n+1}(x^2) - E_n P_{n}(x^2)}{x^2 - \lambda_k},
\end{equation}

\noindent with
\begin{equation}
    E_n = \frac{P_{n+1}(\lambda_k)}{P_{n}(\lambda_k)},
\end{equation}

\noindent and the entries of $\check{A}$ are related to those of $A$ by
\begin{align}
    \check{u}_n &= u_n \frac{E_n}{E_{n-1}},\\
    \check{b}_n &= b_{n+1} + E_{n+1} - E_n.
\end{align}

The transform effectively removes the spectral point $\lambda_k$, hence the name of the technique, and the new matrix $\check{A}$ is of size $N$ instead of $N+1$. However, the positivity of the weight is preserved only if $k=0$ or $k=N$. Removing $\lambda_N$ only amounts to going from $N$ odd to $N$ even or vice versa. Removing $\lambda_0 = 0$ will give rise to a system that is not free-free anymore. Such systems will be discussed in the next section. To remove eigenvalues in the middle of the spectrum, the trick is actually to remove a neighboring pair of spectral points with
\begin{equation}
    \check{w}_s = C (\lambda_s - \lambda_k)(\lambda_s - \lambda_{k+1}) w_s.
    \label{eq:weightSurg2}
\end{equation}

\noindent Since the eigenvalues are alternating in parity, the removal of a consecutive pair of spectral points preserves this characteristic in the remaining set of eigenvalues. The polynomials that are orthogonal with respect to the weight in (\ref{eq:weightSurg2}) are obtained by applying twice the formulas given above. It can also be shown \cite{Vinet_HowTo} that removing such a pair of spectral points preserves persymmetry. One can in fact repeat this removal of a pair of spectral points a number of times to construct a new persymmetric matrix of desired size and spectrum from the matrix $A$ with this property.

\begin{figure*}
    \centering
    \includegraphics[width=0.45\textwidth]{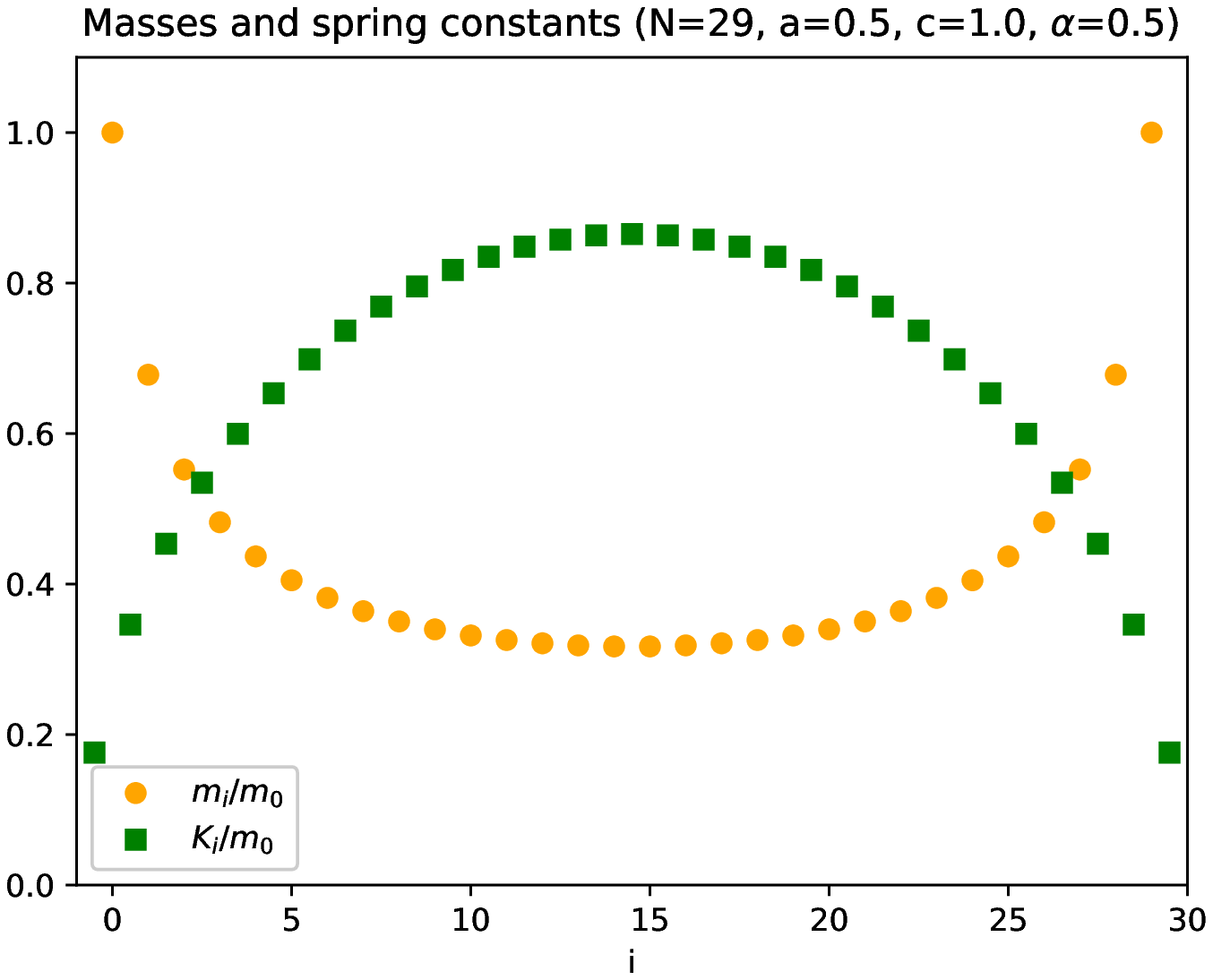}
    \includegraphics[width=0.45\textwidth]{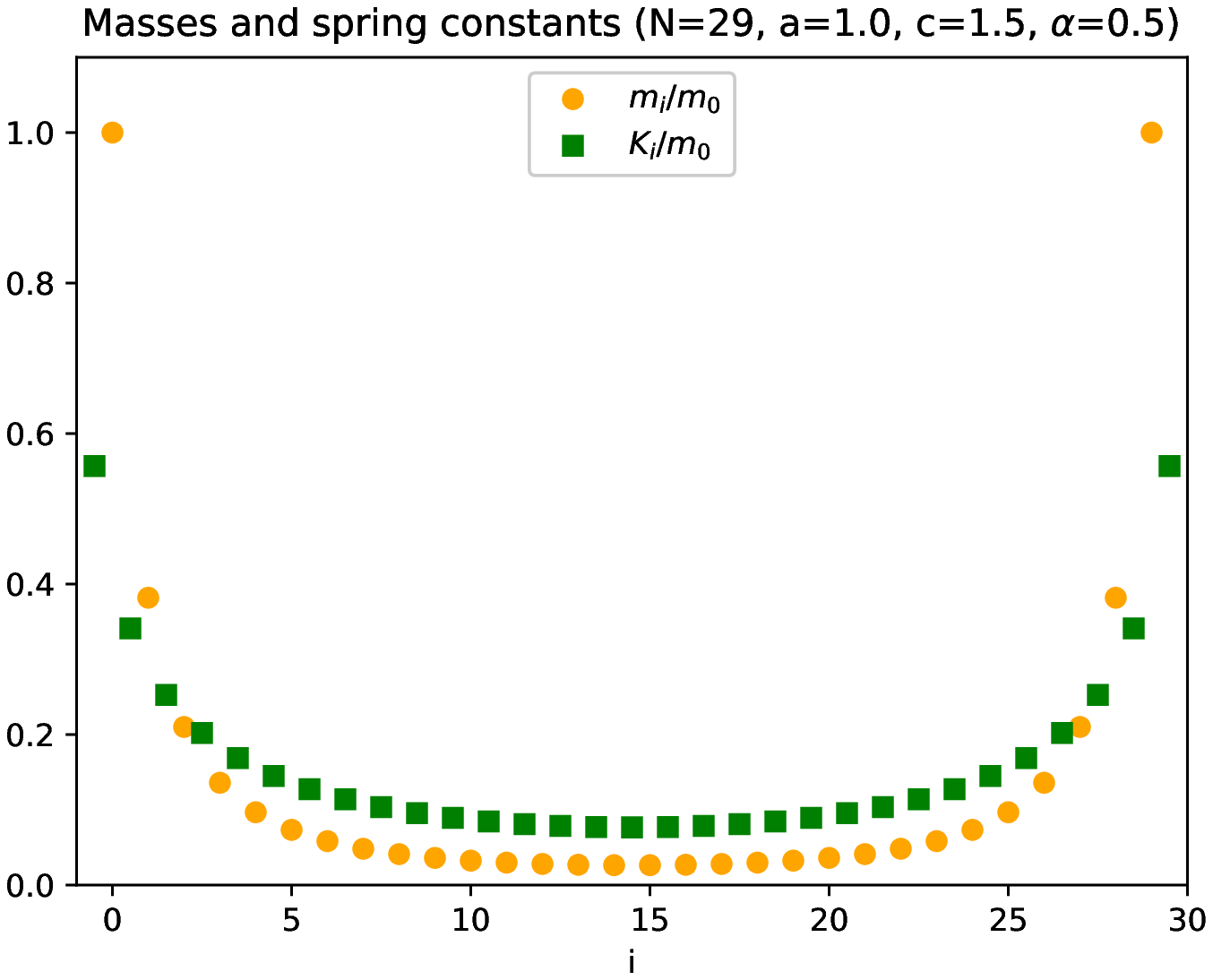}
    \includegraphics[width=0.45\textwidth]{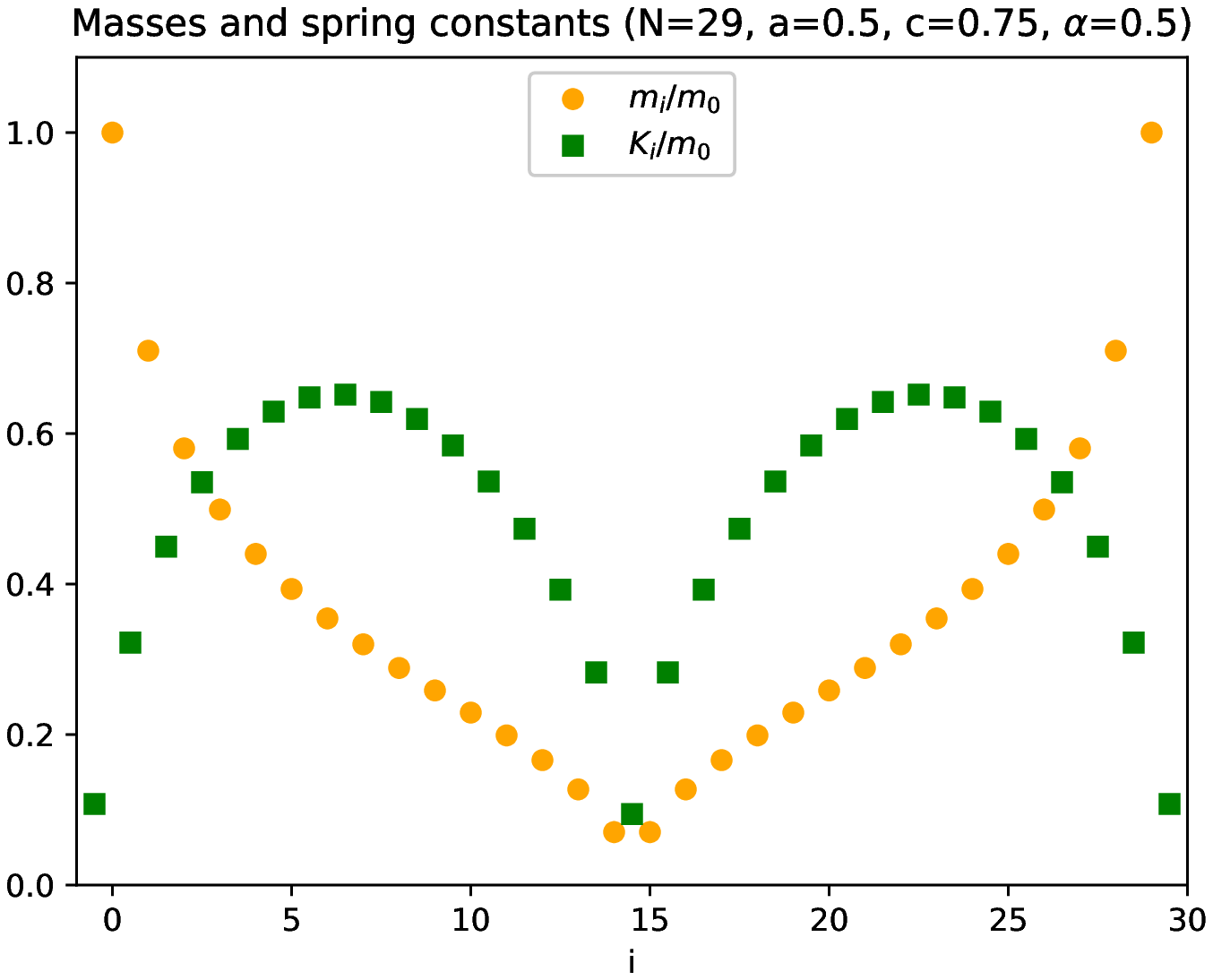}
    \includegraphics[width=0.45\textwidth]{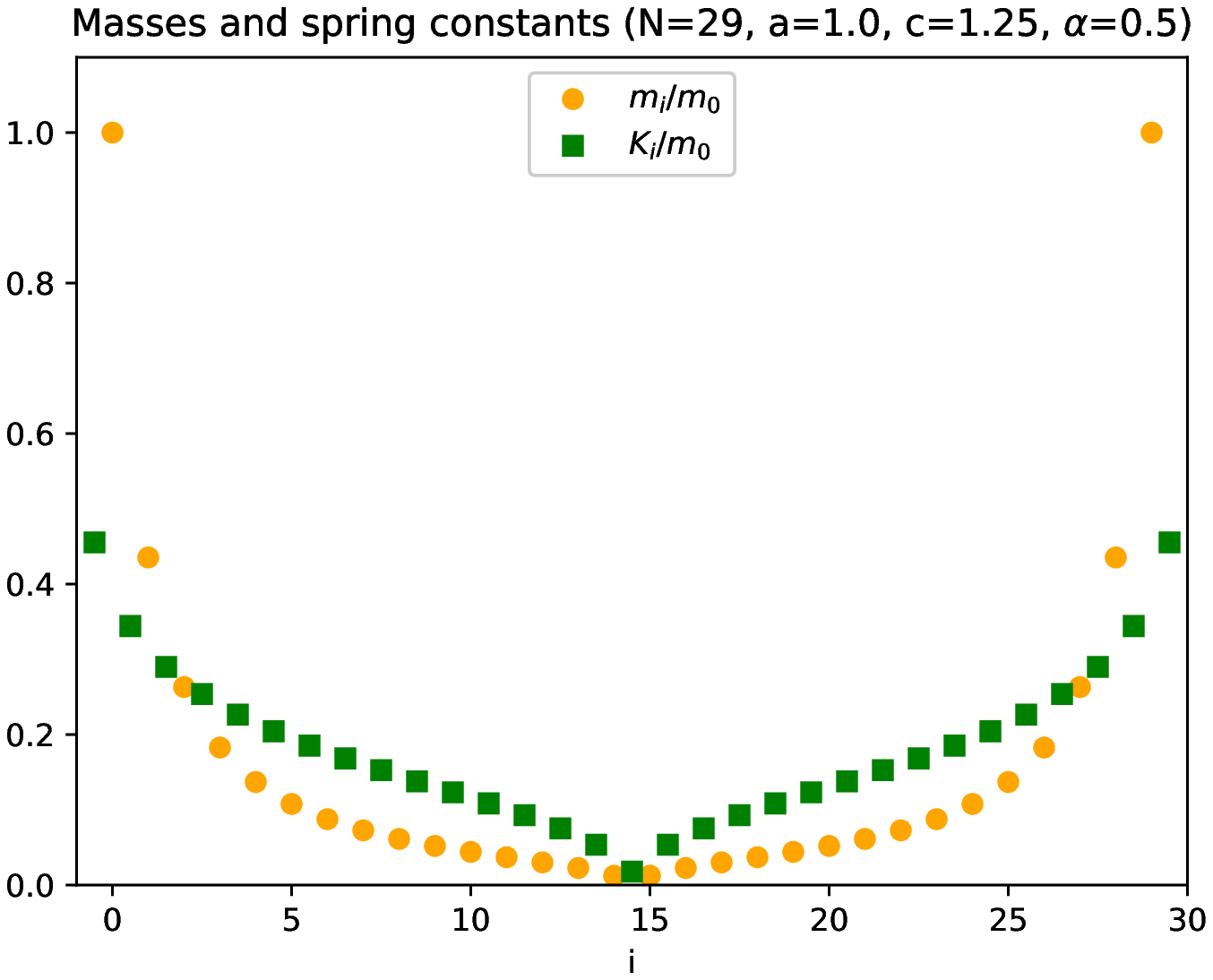}
    \includegraphics[width=0.45\textwidth]{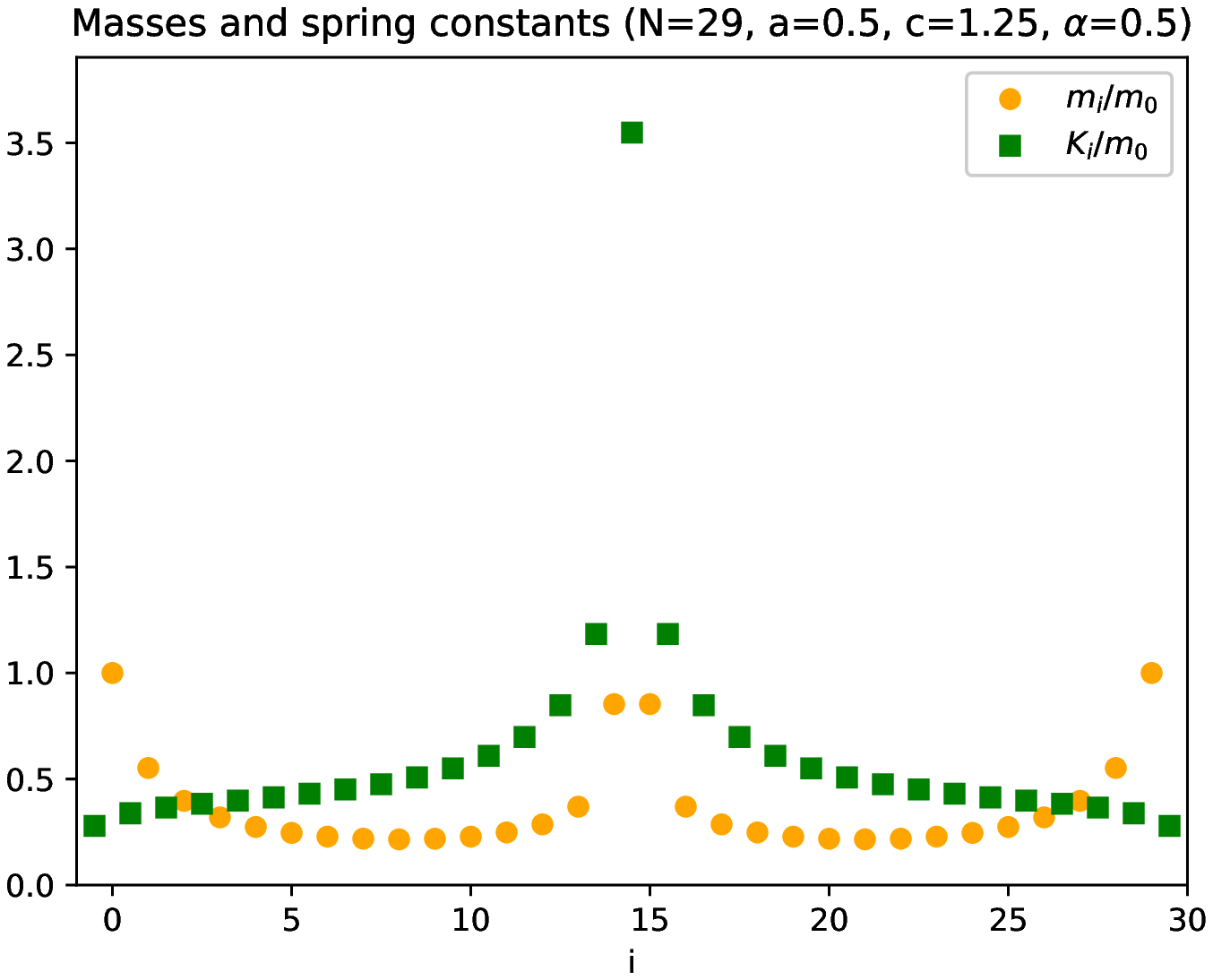}
    \includegraphics[width=0.45\textwidth]{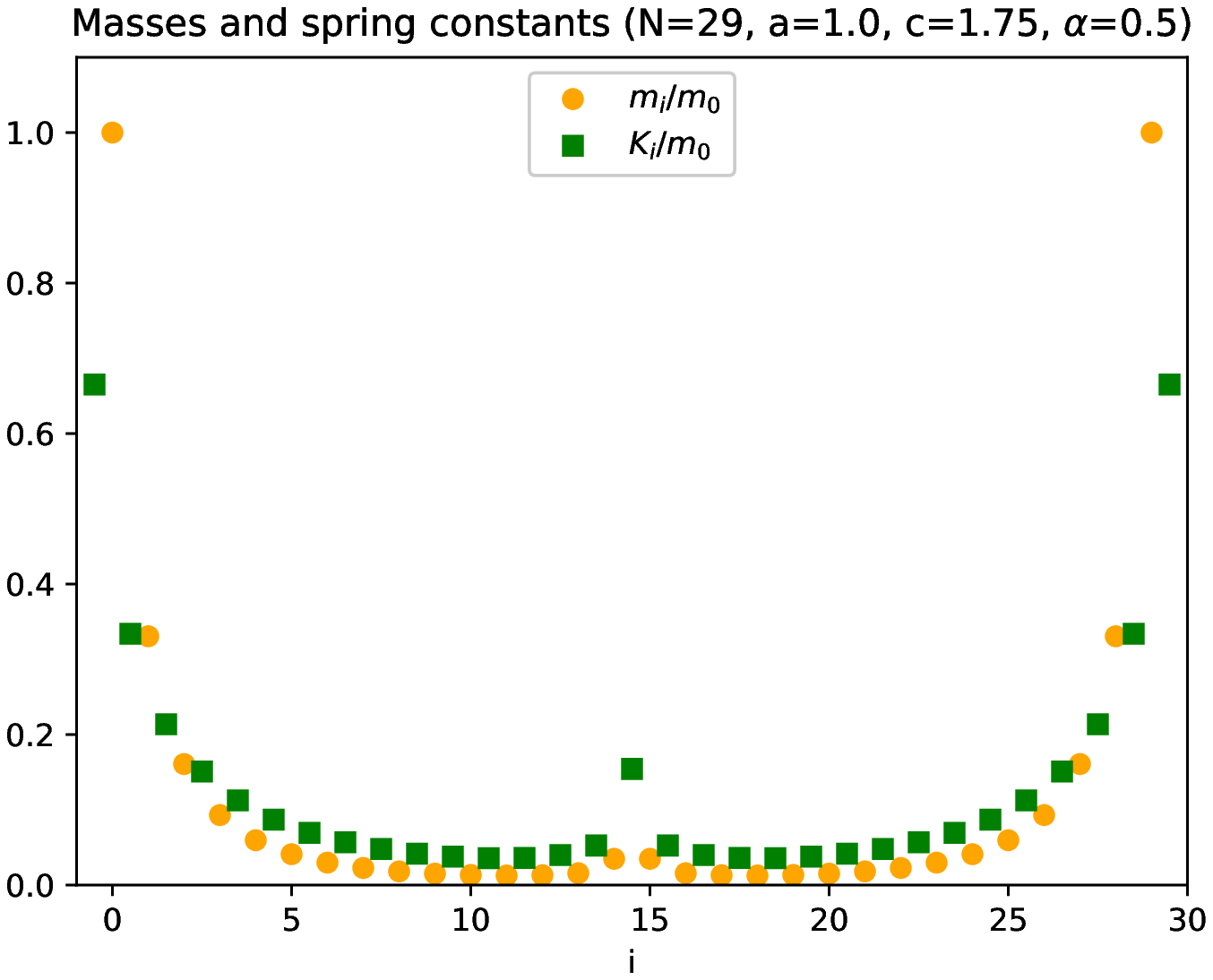}
    \caption{\label{fig:fixed_mK} Normalized masses and spring constants for a mirror-symmetric fixed-fixed system with two different values of $a$ and with $c-a=\frac{1}{2}$, $c-a<\frac{1}{2}$ and $c-a>\frac{1}{2}$ respectively; with $\tilde{\omega}=2\pi/N$ again for adequate visualisation.}
\end{figure*}

\section{Fixed-fixed mass-spring chain}
\label{sec:fixed}

We now look at the fixed-fixed system where the first and last masses are attached with a spring to a wall. We still consider a mirror-symmetric system to be able to find perfect transfer. The Hamiltonian (\ref{eq:HamA}) is essentially the same, and (\ref{eq:mat_A}), (\ref{eq:bi_matrix}) and (\ref{eq:ui_matrix}) remain valid, except that we now have $K_0 = K_{N+1} \neq 0$. The trajectories are very similar. The model is again based on the para-Racah polynomials and so we need $\alpha = \frac{1}{2}$ for persymmetry but we must now have $a \neq 0$ (and $c \neq 0$) since there are no translation mode in this case. The conditions for perfect transfer are also the same, namely $\omega_n = \omega k_n$ with the $k_n$ alternating in parity with $n$. This and the positivity conditions for $u_n$ lead to requiring that
\begin{align}
    a &= \frac{\mu}{Z} > -\frac{1}{2} \qq{with} a \neq 0,
    \label{eq:aFixedFraction}\\
    c &= a + \frac{\rho}{Z} > -a \qq{with} 0 < \frac{\rho}{Z} < 1,
    \label{eq:cFixedFraction}\\
\end{align}

\begin{figure*}
    \centering
    \includegraphics[width=0.45\textwidth]{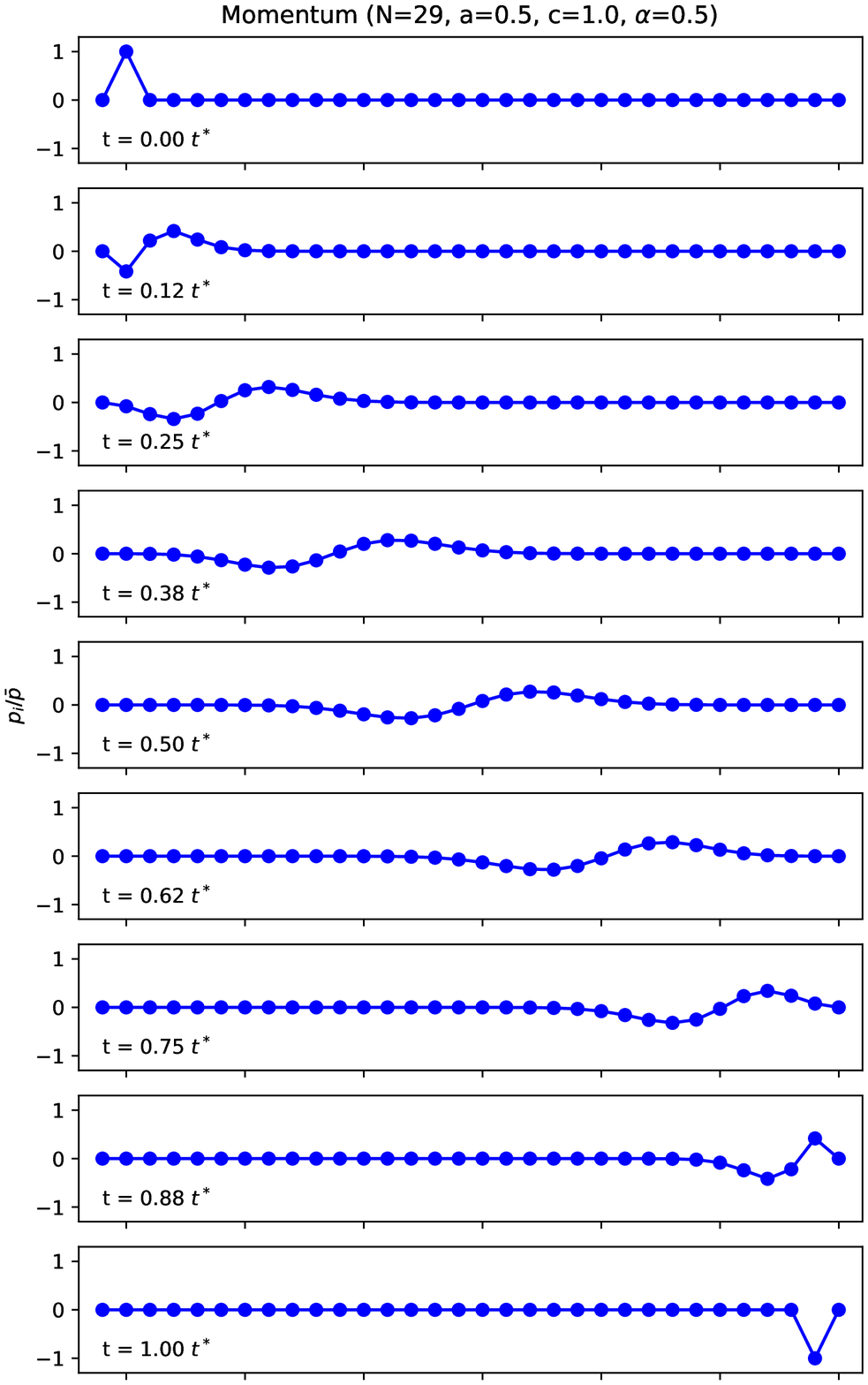}
    \includegraphics[width=0.45\textwidth]{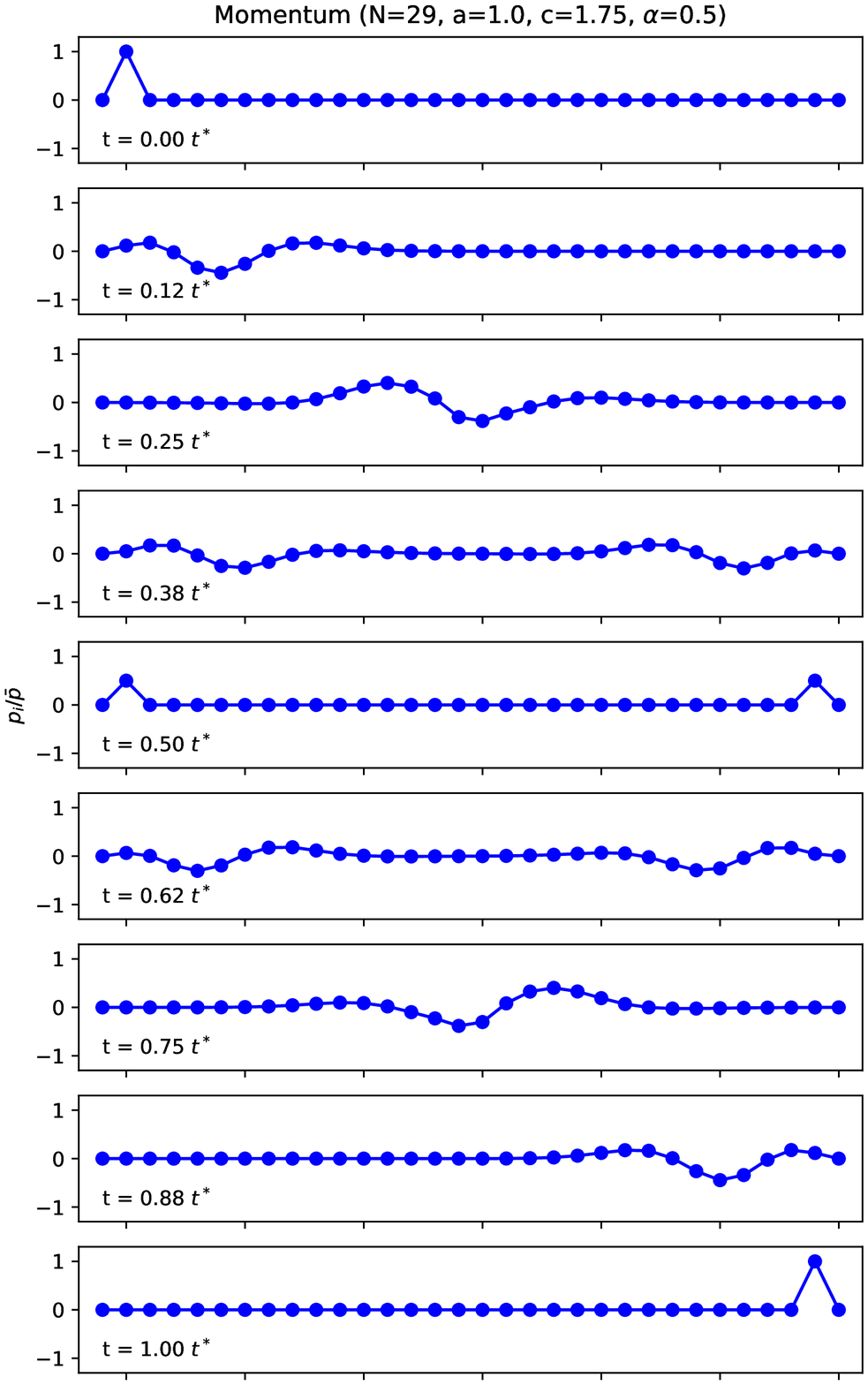}
    \caption{\label{fig:fixed_p_9}Snapshots of the time-evolution of the mass-weighed momentum of a mirror-symmetric fixed-fixed system at various fractions of $t^*$ for $c-a=\frac{1}{2}$ and $c-a \neq \frac{1}{2}$ respectively (video available \cite{suppl}).}
\end{figure*}

\noindent where $\mu$, $\rho$ and $Z$ have no common factor, $Z$ is even and $\rho$ is odd. Again, no generality is lost by choosing a particular set of positivity conditions. The eigenfrequencies are
\begin{align}
    \omega_{2s} &= \frac{\tilde{\omega}}{Z} (Zs + \mu), \quad s = 0,\dots,j,
    \label{eq:w2sFixed}\\
    \omega_{2s+1} &= \frac{\tilde{\omega}}{Z} (Zs + \mu + \rho), \quad s = 0,\dots,j,
    \label{eq:w2s1Fixed}
\end{align}

\noindent for $N$ odd, and the same for $N$ even except that $s$ stops at $j-1$ in (\ref{eq:w2s1Fixed}). Perfect transfer occurs also at $t^* = \pi/\omega$, with $\omega = \tilde{\omega}/Z$. If $\mu$ is even, the momentum of the last mass will be in the same direction as the initial momentum ($p_N(t^*) = \bar{p}$); if $\mu$ is odd, it will be in the opposite direction ($p_N(t^*) = -\bar{p}$), see figure \ref{fig:fixed_p_9} for an example of this behavior.

The entries for $A$ are now given by:
\begin{equation}
    b_n = \frac{\tilde{\omega}^2}{2}\qty[a(a+j)+c(c+j)+n(N-n)],
\end{equation}
\begin{align}
    u_n = \tilde{\omega}^4 \frac{n(N+1-n)(N-n+a+c)(n-1+a+c)}{4(N-2n)(N-2n+2)}\nonumber \\
    \hfill \vdot \qty((n-j-1)^2-(a-c)^2),
\end{align}

\noindent for $N$ odd, and
\begin{align}
    b_n = \tilde{\omega}^2 \frac{(N-n)(n+a+c)(n+a-c-j+1)}{2(2n+1-N)} \nonumber\\
    \hfill + \tilde{\omega}^2 \frac{n(N-n+a+c)(n-j-1+c-a)}{2(2n-1-N)} + \tilde{\omega}^2 a^2,
\end{align}
\begin{align}
    u_n = \tilde{\omega}^4 \frac{n(N+1-n)(a+c+n-1)(a-c+j-n+1)}{4(N-2n+1)^2} \nonumber\\
    \hfill \vdot (c-a+j-n)(a+c+N-n),
\end{align}

\noindent for $N$ even.

The matrix $A$ is no longer singular and can be inverted. Specifically,
\begin{equation}
    A_{ik}^{-1} = \sum_{n=0}^N \frac{1}{\lambda_n} U_{ni} U_{nk}.
\end{equation}

\noindent We can follow the proof of lemma 2 in \cite{Nylen1997} to construct the matrix $M^{-\frac{1}{2}}$. The scaling parameter is now $K_0$. If we define
\begin{equation}
    \gamma_i = \sum_{s=0}^j \frac{1}{\lambda_{2s}} U_{2s,i} U_{2s,0},
\end{equation}

\noindent the normalized expressions for the masses and spring constants are given by
\begin{align}
    \frac{m_i}{K_0} &= \frac{2}{\gamma_0} \gamma_i^2,\\
    \frac{K_i}{K_0} &= \frac{2}{\gamma_0} \gamma_{i-1} \gamma_i \sqrt{u_i},\\
    K_0 &= K_N = (\text{free parameter}),
\end{align}

\noindent which is valid for $N$ odd or even. We can rewrite these expressions to have $m_0$ as the scaling parameter. We find
\begin{align}
    \frac{m_i}{m_0} &= \qty(\frac{\gamma_i}{\gamma_0})^2, \label{eq:mi_fixed}\\
    \frac{K_i}{\tilde{\omega}^2 m_0} &= \qty(\frac{\gamma_{i-1}}{\gamma_0}) \qty(\frac{\gamma_i}{\gamma_0}) \sqrt{\frac{u_i}{\tilde{\omega^4}}}, \label{eq:Ki_fixed}\\
    \frac{K_0}{m_0} &= \frac{K_N}{m_0} = \frac{1}{2 \gamma_0}. \label{eq:K0_fixed}
\end{align}

\begin{figure}
    \includegraphics[width=0.45\textwidth]{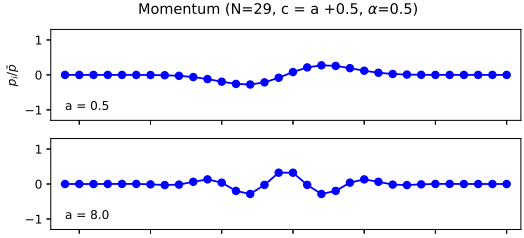}
    \caption{\label{fig:fixed_p_2}Snapshots of the time-evolution of the mass-weighed momentum of a mirror-symmetric fixed-fixed system at $t = \frac{1}{2} t^*$ with smaller $a$ and bigger $a$ respectively (video available \cite{suppl}).}
\end{figure}

\noindent These expressions provide the masses and spring constants of chains of type fixed-fixed with perfect transfer. Notice that, because of (\ref{eq:alternate}), $\gamma_{N-i} = \gamma_i$, and the mirror-symmetry of the chain is confirmed. Examples of values for the masses and spring constants of fixed-fixed chains appear in figure \ref{fig:fixed_mK}. The same trend as in the free-free case can be seen here, with a smooth curve for $c-a = \frac{1}{2}$ and a discontinuity otherwise; moreover the central masses and spring constants tend to become lower or higher if $c-a < \frac{1}{2}$ and $c-a > \frac{1}{2}$ respectively. Also, it can be observed that choosing $a$ bigger will make $K_0$ bigger. In addition, the numerical simulations indicate bigger $a$'s produce waves with more ripples, see for example figure \ref{fig:fixed_p_2}. This is explained by the greater relative value of the boundary spring constant $K_0$, which makes the first mass oscillates many times in the beginning and creates this shape.

\begin{figure*}
    \centering
    \includegraphics[width=0.45\textwidth]{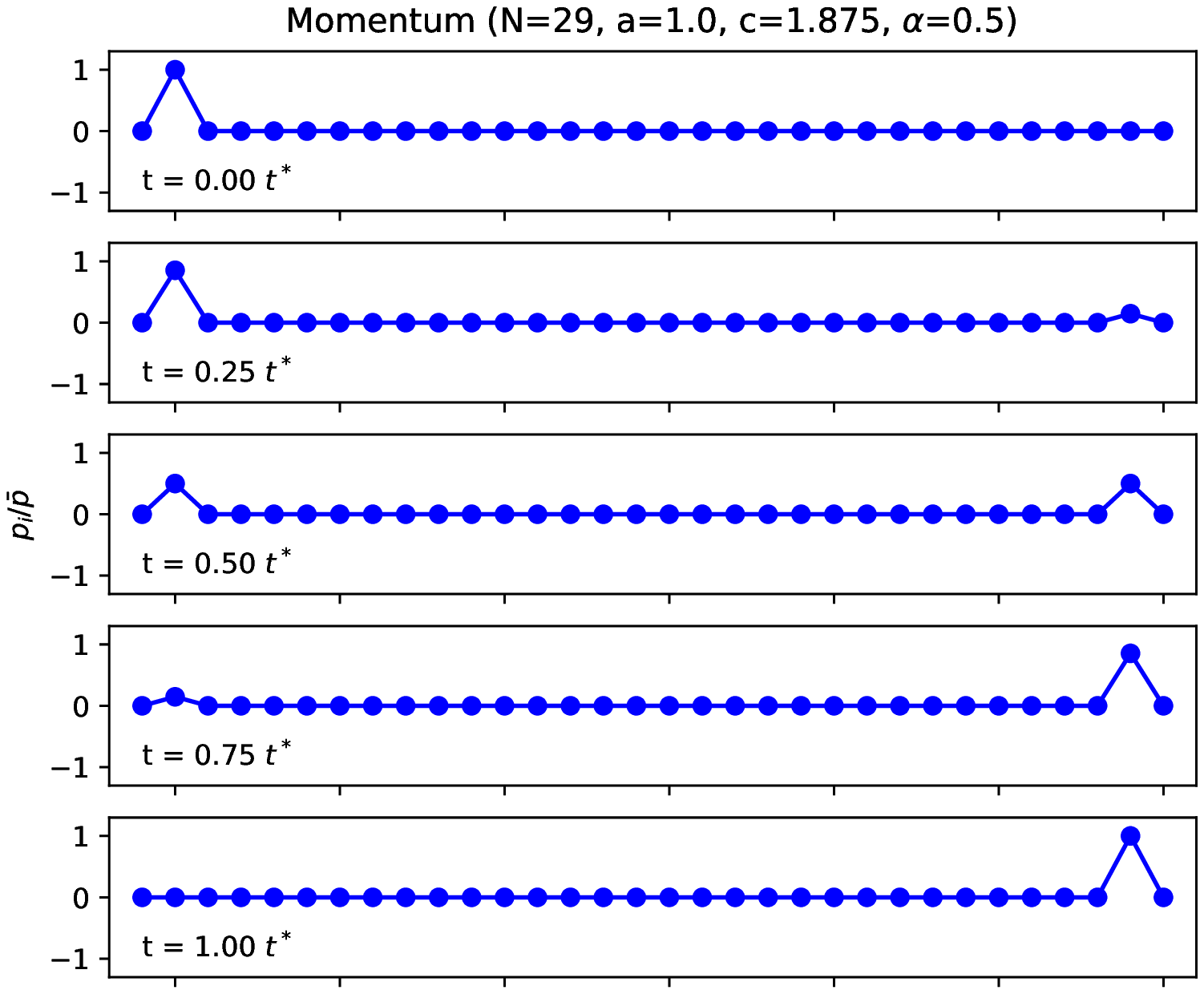}
    \includegraphics[width=0.45\textwidth]{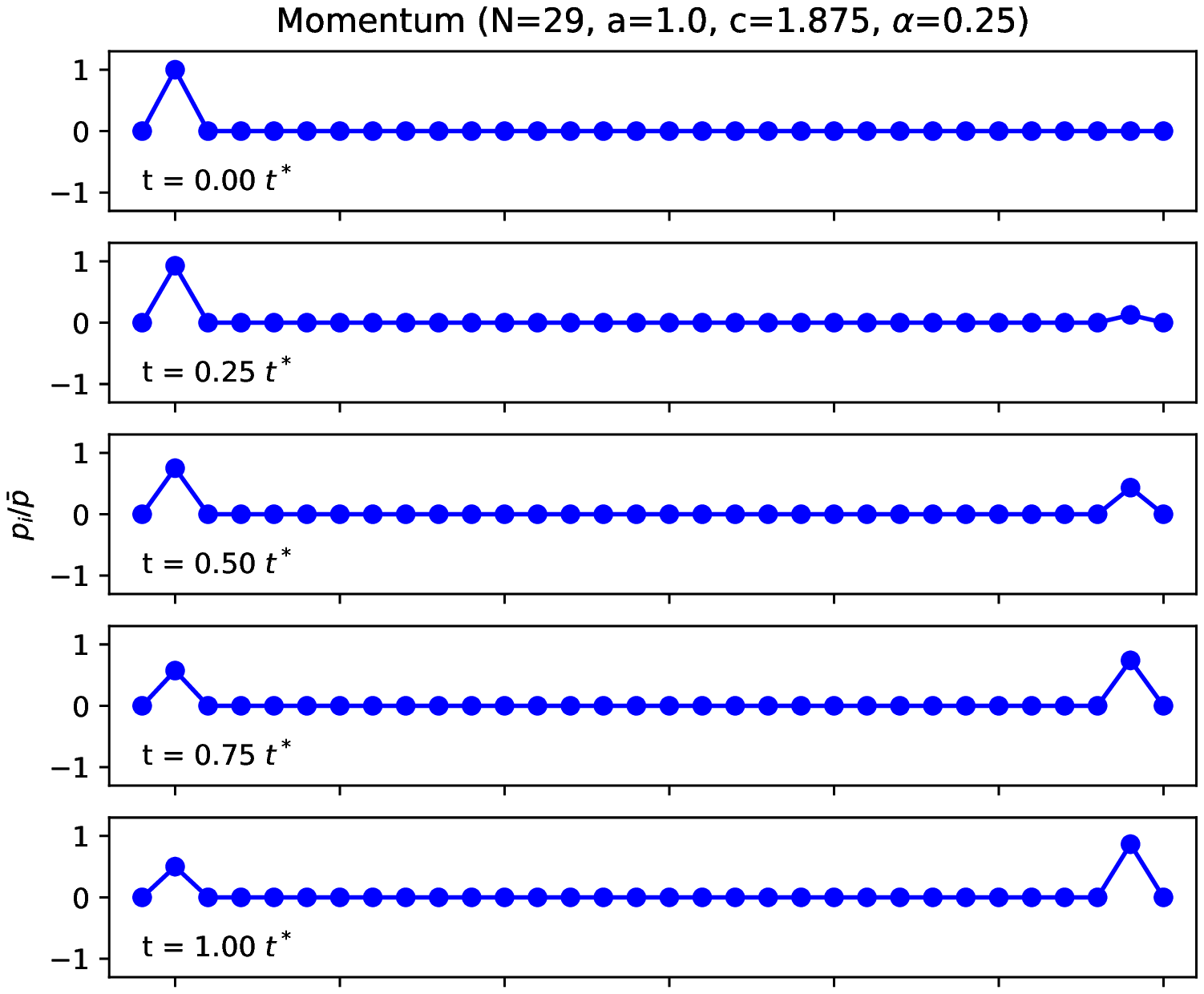}
    \caption{\label{fig:fixed_p_5}Snapshots of the time-evolution of the mass-weighed momentum of a free-free system at various times $\tau_\ell$ exhibiting fractional revival for $\alpha=\frac{1}{2}$ and $\alpha \neq \frac{1}{2}$ respectively; here $Z=8$ (video available \cite{suppl}).}
\end{figure*}

With (\ref{eq:Uni}), and owing to the fact that $P_0 (x^2) = 1$, we can rewrite $\gamma_i$ as
\begin{equation}
    \gamma_i = \sum_{s=0}^j \frac{w_{2s}}{\lambda_{2s}} \frac{P_i(\lambda_{2s})}{\sqrt{u_1 \dots u_i}}. \label{eq:gamma_i}
\end{equation}

We have not found a way to compute this sum. Note that (\ref{eq:b_nFromAC}) yields a different recurrence relation for $y_i$ than for $A_i$ when $a \neq 0$, so the trick used previously to find closed-form expressions does not help here. It is however interesting to see that taking the limit $a \to 0$ leads back to what was found for the free-free case. First, we write
\begin{equation}
    \lim_{a \to 0} \frac{m_i}{m_0} = \lim_{a \to 0} \qty(\frac{\gamma_i}{\gamma_0})^2 = \lim_{a \to 0} \qty(\frac{a^2 \gamma_i}{a^2 \gamma_0})^2,
\end{equation}

\noindent with
\begin{equation}
    a^2 \gamma_i = w_0 \frac{P_i(a^2)}{\sqrt{u_1 \dots u_i}} + a^2 \sum_{s=1}^j \frac{w_{2s}}{\lambda_{2s}} \frac{P_i(\lambda_{2s})}{\sqrt{u_1 \dots u_i}}.
\end{equation}

\noindent Given that $P_i(x^2) = (-1)^i A_0 A_1 \dots A_{i-1} \tilde{P}_i(x^2)$, where $\tilde{P}_i(x^2)$ are the para-Racah polynomials normalized so that $\tilde{P}_i(0) = 1$ when $a=0$ (see \cite{paraRacah}), we have
\begin{align}
    \lim_{a \to 0} \frac{m_i}{m_0}  &= \qty(\frac{P_i(0)}{\sqrt{u_1 \dots u_i}})^2\\
    &= \frac{(A_{i-1} \dots A_0)^2}{u_1 \dots u_i}.
\end{align}

\noindent Noting also that $\lim_{a \to 0} K_0 = \lim_{a \to 0} \frac{1}{2 \gamma_0} = 0$ because of the singularity in $\gamma_0$, this leads to what was derived for the free-free system, which can thus be viewed as the limit when $a$ tends to zero of the more general chain corresponding to the para-Racah polynomials with the parameter $a$ arbitrary. Visually, this tendency is suggested by figure \ref{fig:fixed_mK}, as the left side,  where $a$ is smaller, indeed resembles figure \ref{fig:free_mK}.

It is straightforward to establish that there will be fractional revival at the times $\tau_\ell$ previously defined. First notice that
\begin{align}
    \cos(\omega_{2s} \tau_\ell) &= \cos(2\ell a\pi),
    \label{eq:cos2sfixed}\\
    \cos(\omega_{2s+1} \tau_\ell) &= \cos(2\ell c\pi),
    \label{eq:cos2s1fixed}
\end{align}

\noindent which do not depend on $n$ (or $s$). From that and (\ref{eq:sumU2s}) and (\ref{eq:sumU2s1}), the fractional revival is described by
\begin{eqnarray}
    \frac{p_i(\tau_\ell)}{\bar{p}} &=& \delta_{i0} \cos((c+a)\ell \pi)\cos((c-a)\ell \pi) \nonumber\\
    & & + \delta_{iN} \sin((c+a)\ell \pi) \sin((c-a)\ell \pi)
    \label{eq:pFRfixedAlpha}
\end{eqnarray}

\noindent which reduce to (\ref{eq:pFRfree}) if $a=0$. The behavior of the pulse is similar to that of the free-free case, with the wave ``hitting'' the discontinuity in the middle of the chain and generating fractional revival, this time only if $c - a \neq \frac{1}{2}$, see figure \ref{fig:fixed_p_9}. If one is only interested in fractional revival, the conditions on the parity of $\mu$, $\rho$ and $Z$ can be relaxed. In fact, $a$ and $c$ can be taken to be real parameters as long as $a > -\frac{1}{2}$ and $\abs{a} < c < \abs{a+1}$, but again the system will return to its initial state only if $a$ and $c$ are rational.

The isospectral deformation also gives a system with fractional revival thar correspond to the para-Racah polynomials with the parameter $\alpha$ arbitrary. The system is completely determined once we choose $\tilde{K}_0$ and $\tilde{K}_{N+1}$ \cite{Nylen1997}. Then, if we choose $\tilde{K}_0 = K_0$ and $\tilde{K}_{N+1} = \frac{\alpha}{1 - \alpha} \tilde{K}_0$, it is straightforward to deduce that the relations (\ref{eq:mi_alpha_odd}) to (\ref{eq:Ki_alpha_even}) still hold for the fixed-fixed system. This choice of parameters for $\tilde{K}_0$ and $\tilde{K}_{N+1}$ is the only one that will lead to conservation of momentum at times $\tau_\ell$ for this system, since this quantity is not necessarily conserved when the system is fixed. Fractional revival is encapsulated in
\begin{eqnarray}
    \frac{p_i(\tau_\ell)}{\bar{p}} &= \delta_{i0} \qty[ (1-\alpha)\cos(2 \ell a \pi) + \alpha \cos(2\ell c \pi)] \nonumber\\
    & + \delta_{iN} \sqrt{\alpha (1-\alpha)}\qty[ \cos(2 \ell a \pi) - \cos(2\ell c \pi)]
    \label{eq:pFRfixed}
\end{eqnarray}

\noindent which coincides with (\ref{eq:pFRfreeAlpha}) if $a=0$. Again, perfect transfer is only possible if $\alpha = \frac{1}{2}$, with (\ref{eq:pFRfixed}) then reducing to (\ref{eq:pFRfixedAlpha}), see for example figure \ref{fig:fixed_p_5}. In summary, in the fixed-fixed case, the system exhibits perfect transfer without fractional revival for $\alpha = c-a = \frac{1}{2}$, perfect transfer and fractional revival for $\alpha = \frac{1}{2}$ and $c-a \neq \frac{1}{2}$ with $a$ and $c$ defined as in (\ref{eq:aFixedFraction}) and (\ref{eq:cFixedFraction}), and fractional revival only for other choices of $a$, of $c$, or of $\alpha$. Finally, the spectral surgery can also be performed in the fixed-fixed case.

\section{Conclusion}
\label{sec:conclusion}

Summing up, we have elaborated on the possibility of constructing classical analogs of quantum systems with perfect state transfer and fractional revival. Indeed, analytic mass-spring chains with similar properties have been obtained. The additional difficulty in the classical case is to identify Jacobi matrices whose eigenvalues are perfect squares. These were provided by the quadratic bi-lattice formed by the spectral points of the para-Racah polynomials. Perfect transfer also heavily relies on the mirror-symmetry of the chain, or the persymmetry of the matrix diagonalized by the polynomials, and this happens for a special case of the para-Racah polynomials. It would of course be of interest to examine beyond the case treated here, if other families of orthogonal polynomials could provide analytic Newton's cradles.

We have concentrated on the transmission of a pulse from the first mass to the last. It is worth pointing out that, because of its mirror-symmetry, the system will actually reproduce at time $t=t^*$ the mirror image of whatever its initial condition was at time $t=0$, i.e.
\begin{equation}
    p_i(0) = \bar{p}_i \implies p_i(t^*) = \bar{p}_{N-i}.
\end{equation}

\noindent Indeed, one could transmit to mass $m_{N-i}$ a pulse given to mass $m_i$, or even send more complicated signals, for example a 2-bit message from masses $m_0$ and $m_1$ to masses $m_N$ and $m_{N-1}$. Such a system could even be used to transmit a signal from the first few masses to their mirror images at the other end, while at the same time transmitting a signal encoded in masses near the end to masses at the beginning in view of the perfect reversal of the initial condition at time $t^*$.

The last possible type of mass-spring chains is one where the first mass is fixed, but the last mass is free to move, i.e. where $K_0 \neq 0$ and $K_{N+1} = 0$ (or vice versa). Such a chain was not discussed so far because it cannot be mirror-symmetric, and hence will not lead to perfect transfer between the extremities of the chain. Indeed, (\ref{eq:PST_final}) is valid for the perfect transfer of the pulse if and only if the first and last mass are the same. However, we still can reconstruct $M$ and $K$ of type fixed-free from the non-singular persymmetric matrix $A$ of section \ref{sec:fixed} using the computations described in lemma 1 of \cite{Nylen1997}, and the behavior will be similar to that of the fixed-fixed persymmetric system, given the initial conditions (\ref{eq:init_cond}). Indeed, everything presented in the mass-weighed coordinates in the fixed-fixed section would still apply, but the actual pulse will not be completely transmitted. Perfect state transfer without mirror-symmetry has been shown to be possible in quantum spin chains between sources and targets that are asymmetrically located \cite{Coutinho_2019, Kay_2011}. It would be interesting to investigate such behavior in a fixed-free mass-spring chain, as well as in non-mirror-symmetric free-free and fixed-fixed systems. More generally, it should be quite instructive to continue dwelling into the classical correspondence of the large body of knowledge that has been developed on perfect state transfer in quantum frameworks.

\begin{acknowledgments}
H.S. benefitted from a Undergraduate Student Research Awards (USRA) scholarship from the Natural Sciences and Engineering Research Council of Canada (NSERC). The researchof L.V. is supported in part by a Discovery Grant from NSERC. The work of A.Z. is funded by the National Science Foundation of China (Grant No.11771015). A.Z. gratefully acknowledges the hospitality of the CRM over an extended period and the award of a Simons CRM professorship.
\end{acknowledgments}



%

\end{document}